\documentclass[iop]{emulateapj}
\usepackage{graphicx}
\usepackage{epsfig}
\usepackage{amsmath,amssymb}
\usepackage{verbatim}
\usepackage{subfigure}
\usepackage{epstopdf}
\usepackage{chngcntr}

\def\lapprox{\lower.4ex\hbox{$\;\buildrel <\over{\scriptstyle\sim}\;$}}
\def\gapprox{\lower.4ex\hbox{$\;\buildrel >\over{\scriptstyle\sim}\;$}}
\def\na{\ref@jnl{New A}}                % New Astronomy

\shorttitle{Modeling X-ray Emission}
\shortauthors{Anderson et al.}

\begin{document}
\title{Modeling X-ray Emission Around Galaxies}
\author{Michael E. Anderson\altaffilmark{1,2}, Joel N. Bregman\altaffilmark{1}}
\altaffiltext{1}{Department of Astronomy, University of Michigan, Ann Arbor, MI 48109}
\altaffiltext{2}{Max-Planck Institut f\"ur Astrophysik, Garching bei M\"unchen, Germany; michevan@mpa-garching.mpg.de}

\begin{abstract}
Extended X-ray emission can be studied either spatially (through its surface brightness profile) or spectrally (by analyzing the spectrum at various locations in the field). Both techniques have advantages and disadvantages, and when the emission becomes particularly faint and/or extended, the two methods can disagree. We argue that an ideal approach would be to model the events file directly, and therefore to use both the spectral and spatial information which are simultaneously available for each event. In this work  we propose a first step in this direction, introducing a method for spatial analysis which can be extended to leverage spectral information simultaneously. We construct a model for the entire X-ray image in a given energy band, and generate a likelihood function to compare the model to the data. A critical goal of this modeling is disentangling vignetted and unvignetted backgrounds through their different spatial distributions. Employing either maximum likelihood or Markov Chain Monte Carlo, we can derive probability distribution functions for the source and background parameters together, or we can fit and subtract the background, leaving the description of the source non-parametric. We calibrate and demonstrate this method against a variety of simulated images, and then apply it to Chandra observations of the hot gaseous halo around the elliptical galaxy NGC 720. We are able to follow the X-ray emission below a tenth of the background, and to infer a hot gas mass within 35 kpc of $4-5\times10^9 M_{\odot}$, with some indication that the profile continues  to at least 50 kpc and that it steepens as the radius increases. We derive much stronger constraints on the surface brightness profile than previous studies, which employed the spectral method, and we show that the density profiles inferred from these studies are in conflict with the observed surface brightness profile. We therefore conclude that, contrary to a previous claim, the X-ray halo of NGC 720 does not seem to contain the full complement of missing baryons from this galaxy. \\
\end{abstract}


\maketitle

\section{Introduction}

The study of very extended emission comprises a large portion of the work of extragalactic X-ray astronomy. All galaxy clusters (\citealt{Gursky1977}, \citealt{Forman1982}, \citealt{Rosati2002}) and most galaxy groups (\citealt{Mulchaey1993}, \citealt{Ebeling1994}, \citealt{Ponman1996}, \citealt{Mulchaey2000}) are suffused with a hot ($kT > 10^6$ K), X-ray emitting gaseous medium. In all clusters and many groups, this medium contains the majority of the baryons in the system (\citealt{Ettori2003}, \citealt{Gonzalez2007}, \citealt{Giodini2009}, \citealt{Andreon2010}, \citealt{Dai2010}, \citealt{Sanderson2013}), and extends to hundreds of kpc. In recent years, X-ray observations have even been able to push outwards to the virial radius of some nearby clusters (e.g. \citealt{George2009}, \citealt{Bautz2009},  \citealt{Reiprich2009}, \citealt{Hoshino2010}, \citealt{Kawaharada2010}, \citealt{Simionescu2011}, \citealt{Akamatsu2011}, \citealt{Bonamente2012}, \citealt{Morandi2012}, \citealt{Sato2012},  \citealt{Walker2012a}, \citealt{Walker2012b}, \citealt{Ichikawa2013}, \citealt{Walker2013}, \citealt{Urban2014}).

Individual galaxies are also surrounded by extended X-ray emitting halos. Around elliptical galaxies, these hot gaseous halos have been studied for decades (e.g. \citealt{Forman1979}, \citealt{Forman1985}, \citealt{Fabbiano1989}, \citealt{Mathews2003}, \citealt{O'sullivan2001}, \citealt{Mulchaey2010}). Starbursting spirals have extended coronae above and below the disk extending to a few tens of kpc (\citealt{Strickland2004a}, \citealt{Li2006}, \citealt{Tullmann2006}, \citealt{Owen2009}, \citealt{Yamasaki2009}, \citealt{Li2013}). We also recently reported the detection of hot gaseous halos around more quiescent massive spiral galaxies, extending out to $\sim 50$ kpc (\citealt{Anderson2011}, \citealt{Dai2012}, \citealt{Anderson2013}); \citet{Bogdan2013} have confirmed one of these detections and have discovered another hot halo as well. Very extended hot gas is also detected around merging galaxies such as NGC 6240 \citep{Nardini2013}. 

As both of these fields continue to detect emission at larger radii and lower X-ray surface brightness, it is becoming increasingly important to have effective observational techniques for studying faint, extended X-ray emission. At present, there are two major approaches to this analysis: spectral fitting and spatial binning. Spatial binning is conceptually simple: one measures the X-ray radial surface brightness profile in a given band, and infers a gas density profile from the surface brightness profile. The major uncertainties with this method are flat-fielding the image and estimating the background. For bright sources, blank-sky backgrounds are sufficient, but for faint emission the background should be estimated in-field, which requires accurate flat-fielding. A secondary concern is separating various components of emission, if they exist; examples of this separation are illustrated in \citet{Anderson2011}, \citet{Dai2012}, and \citet{Bogdan2013}. 

For the spectral method, one measures the X-ray spectrum in various regions, instead of the surface brightness profile. Analyzing a spectrum requires more photons than measuring a broad-band surface brightness, in part because most realistic spectra have many more free parameters than a surface brightness profile. Thus, the image is usually broken into large radial annuli, which sacrifices some location information. On the other hand, the various instrumental and background components are included in the spectral model, so in theory this method does not require separate flat-fielding or background subtraction. Also additional source components can be included, obviating concerns about confusion between hot gas emission and other X-ray sources such as X-ray binaries or background point sources. The primary downsides to this method are model specification and the need for more photons. 

The aforementioned issues with these methods can be quite important, and can lead to conflicting results. For the spatial method, one major failure mode in flat-fielding is incorrect estimation of the ``vignetting'' - the decrease in sensitivity of the detectors as off-axis angle increases. The vignetting profile varies as a function of time (typically getting worse as the telescope degrades), and must be computed separately for each observation. It also varies as a function of energy, so a given exposure map (which contains the information about the vignetting) should only be computed for and applied to a narrow energy band over which the vignetting effects do not vary significantly. An example of this difficulty occurs in \citet{Pedersen2006}, where the authors found evidence for a hot halo around the giant spiral NGC 5746 at $4\sigma$ significance. Later, after applying an updated calibration file which accounted more correctly for the time-dependent degradation of the instruments, the vignetting profile changed and the signal disappeared \citep{Rasmussen2009}. 

For the spectral method, model specification is particularly important. In the observations of interest today, the signal from hot gas is lower than the background, so the model for the background components in the spectrum can significantly influence the inferred properties of the signal. It is not trivial to construct a model for the X-ray background, and most studies use slightly different prescriptions. The components generally include the extragalactic AGN background at hard energies, and Solar wind X-rays (time variable), emission from the Local Hot Bubble, and the Galactic hot halo at softer energies. These components are all variable (spatially and sometimes temporally) and so their normalizations are not typically known a priori. Moreover, the emission itself can be quite spectrally complex - especially for the Local Bubble and Solar wind X-rays, where charge exchange can affect the signal and the gas is not necessarily in collisional equilibrium. Without sufficient photons to fit all of these spectral components, this method is very susceptible to systematic errors, either from degeneracies between free model parameters, or as a result of fitting the data with an inappropriate model and/or overfitting the data.

An example of this issue can be seen with the isolated elliptical galaxy NGC 720. This galaxy has a massive hot halo that has been observed once with Suzaku, twice with XMM-Newton, and several times with Chandra. One of the Chandra observations was studied by \citet{Humphrey2006}, who used the spectral method in eight annuli to measure the hot gas density profile out to 90 kpc; extrapolating their density profile to 300 kpc yields a hot halo mass of $1\times10^{11} M_{\odot}$, which implies that the galaxy is missing about 1/2 of its expected cosmological allotment of baryons. The rest of the data is analyzed in \citet{Humphrey2011}, with slight changes in their model, yielding a hot halo mass of $3\times10^{11} M_{\odot}$ within $R_{200}$, which implies that the galaxy is baryon complete. Based on the statistical errors quoted in these papers, this is about a $3\sigma$ discrepancy in the mass. About half the discrepancy is caused by the addition of the Suzaku data, and half by the changes in their modeling; however, at a smaller radius like $100$ kpc, the discrepancy is still nearly $3\sigma$, and this difference is caused almost entirely by their modeling. We will discuss this galaxy in much more detail in section 5.

Finally, there are also situations where the spectral method and the spatial method yield different conclusions. A notable example is in the estimation of galaxy cluster density profiles near the virial radius - where the cluster emission is much fainter and the systematic uncertainties in these methods become more important. There has been some debate over the putative detection of a flattening in the radial decrease of the hot gas density profile, and this debate seems largely to fall along the lines between these two methods. The flattening was first observed with Suzaku, using spectral methods (\citealt{George2009}, \citealt{Simionescu2011}); spatial methods, using ROSAT, have often not confirmed this result (\citealt{Ettori2009}, \citealt{Eckert2011}, \citealt{Eckert2012}). Thus, while seems to be general consensus about a number of other features in the gas properties near the virial radius (such as decreasing temperature and flattening entropy), more work needs to be done to understand the behavior of the gas surface brightness and density (and therefore the total gas mass and baryon fraction of the cluster). Relatedly, discrepancies at the 5\%-15\% level for derived parameters such as luminosity, temperature, and pressure have also been noted by \citet{Rozo2012} in samples of clusters analyzed with different techniques (including combinations of spectral and spatial). 

In this paper we explore a potential improvement to the spatial method, taking into account both vignetted and unvignetted backgrounds based entirely on in-field data. This approach is similar to the use of the off field-of-view (OFOV) events for XMM-Newton, but can also be used for Chandra and Suzaku imaging, for which these events are not generally available. Adding an unvignetted component to the background model allows the entire image to be flat-fielded simultaneously, which is important for precise measurements of faint signals. 

The rest of the paper is devoted to taking the first steps towards full image modeling for X-ray astronomy. X-ray observations possess an unusual advantage over other wavelengths in that one records time, position, and energy measurements for each event. Most of this information is discarded when producing an image or a spectrum from an events file, but we argue that computational power has evolved to the point where it is no longer necessary to discard this information. We think the ultimate goal, which is outside the scope of this paper, is to study the X-ray events file directly instead of binning it spatially or spectrally to produce an image or a spectrum. With a good model for the events file that includes both energy and position (and potentially time), one could combine the best features of both spatial and spectral analysis, while minimizing the systematic errors associated with each. 

Here we implement a much more limited form of image modeling, using only spatial information in one energy band (typically 0.5-2.0 keV). We construct a simple model with both vignetted and unvignetted backgrounds.  We then discuss the likelihood function for the image, and compare several different forms for the likelihood function before settling on one function. We show that this likelihood function is able to recover the input parameters in a variety of simulated images. Finally, we apply the method to the case of NGC 720 and show that even this limited form of image modeling offers significant advantages over traditional spectral fitting. Errors are quoted at $1\sigma$ unless otherwise noted.

\section{The Unvignetted Background}

In an infinitely long X-ray observation, most of the hard cosmic X-ray background could be resolved into individual AGNs, with a $\sim 10\%$ contribution from the intracluster medium (\citealt{Comastri1995}, \citealt{Ueda2003}, \citealt{Bauer2004}, \citealt{Hickox2007a}, \citealt{Hickox2007b}). Below about 2 keV, star-forming galaxies begin to contribute background X-ray point sources as well, also at about the 10\%-20\% level. There is also a diffuse X-ray component to the soft emission, due to the local hot ISM (i.e. the Local Hot Bubble) and the Galactic hot halo (\citealt{Snowden1990}, \citealt{Snowden1998}). These various components are all spatially variable across the sky, but in practice this is not as much of a concern as one might expect. The faint point sources, which are not resolved in typical X-ray observations, are so numerous as to wash out most statistical fluctuations in a typical field of view. The bright point sources are treated separately in our model (section 3.3). And the diffuse soft emission typically shows features on large (i.e. degree) scales, which are not very important in a single field of view, although it is still unclear  to what extent the diffuse emission also contains substructure on smaller scales (\citealt{Soltan2005}, \citealt{Galeazzi2009}). All of these backgrounds are focused by the telescope's mirrors, which introduces a vignetting effect across the image: emission that falls at larger radii from the aimpoint is slightly attenuated compared to emission right on the aimpoint. The magnitude of the attenuation is energy dependent; in the soft band the attenuation reduces the signal by roughly a quarter to a third at large off-axis angles. 

There is another type of background which is not focused by the telescope optics. The particle background -- mostly soft protons from the Solar Wind -- impinges directly on the X-ray detectors, and also triggers X-ray fluorescence from the instruments themselves. This background also registers as events on the detectors. Many of the cosmic rays are automatically flagged by the standard data reduction scripts, and during periods of especially high particle flux the entire event file is generally excluded as well. However, some particles cannot be automatically distinguished from X-rays and remain in the image as false positives; instrumental X-rays are also not automatically removed and will appear in the events file. Since this background is not focused by the optics, it does not follow the same vignetting profile as the cosmic and Galactic X-ray backgrounds. 

The unfocused background is not appreciably vignetted at all. In Figure 1, we present an image of the stowed background in the 0.5-2.0 keV band for Chandra ACIS, taken from the most recent observations of the stowed background\footnote{Events file produced by M. Markevitch, available at http://cxc.harvard.edu/contrib/maxim/acisbg}. The background is very nearly uniform across all the frontside-illuminated ACIS-I chips. On the ACIS-S chips there is a top-to-bottom gradient which may be due to charge transfer inefficiency in the readout from these chips and seems to have a fairly persistent spatial distribution. For XMM-Newton EPIC images, the instrumental background is distributed significantly differently for each instrumental emission line, and standard background images have been created which model the overall unfocused background (\citealt{Lumb2002}, \citealt{Kuntz2008},
\citealt{Snowden2013}). The broad-band unfocused background is largely uniform, however.

\begin{figure*}
\begin{center}
\includegraphics[width=12cm]{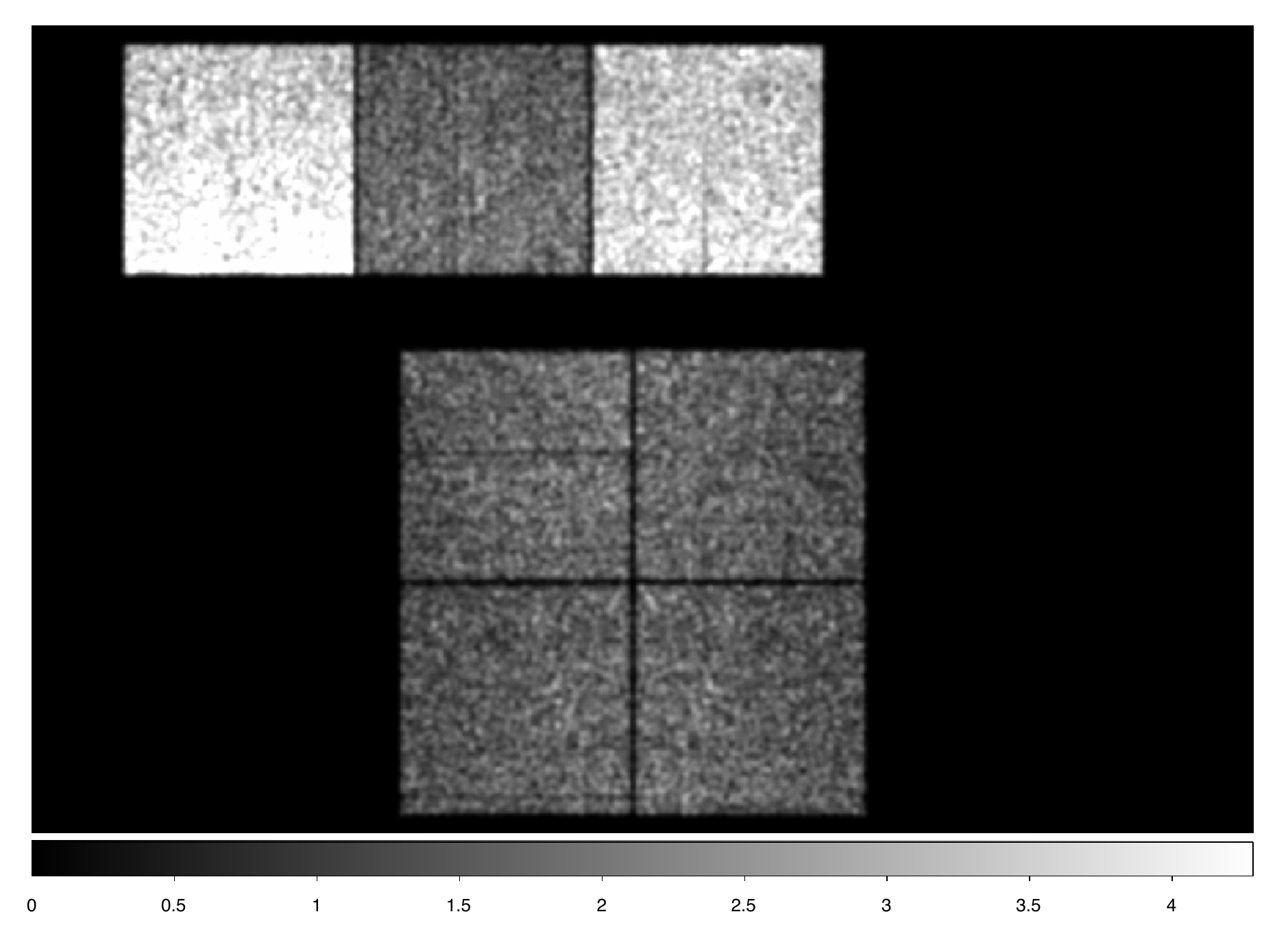}
\caption{0.5-2.0 keV image of the most recent (period E) study of the Chandra stowed background, which should be similar to the overall unfocused background. The image has been smoothed with a Gaussian kernel of radius 3 pixels. The background is very nearly uniform across all the ACIS-I chips. There is a gradient visible across the ACIS-S chips, of order 20-30\%, whose shape does not seem to vary significantly with time. In our analysis, when examining the ACIS-S chips we construct a model for the unvignetted Chandra background based on a smoothed version of this image.}
\end{center}
\end{figure*} 

The spectral shape of the unfocused background is quite complex, and somewhat variable. In the soft bands the flux is much higher at $E \lapprox 0.5$ keV, and otherwise is dominated by prominent emission lines from Aluminum, Silicon, and Gold. In the hard bands the spectrum is fairly flat from 3-5 keV, and then turns upwards and begins to show a number of emission lines from various elements in the detectors. An excellent comparison of the instrumental spectra of Chandra, XMM-Newton, and Suzaku can be seen in Figure 8.2 of \citet{Arnaud2012}. 

With XMM-Newton, the two backgrounds can be disentangled through use of the unexposed off-field of view (OFOV) events, which are by definition entirely composed of unfocused background. This can be used to set the normalization for this component, and then one can apply standard maps of the unfocused background to an observation in order to account for this component. For Chandra and Suzaku, there is currently no straightforward way to separate the two backgrounds. Regardless of instrument, the general practice in reducing X-ray data is to remove as much of this background as possible\footnote{As an aside, we note that the presence of OFOV data is extremely useful in modeling the background, and it might be worthwhile to include these regions as a design consideration in future X-ray missions.}. Filtering periods of the observation with higher count rates reduces the particle background significantly, as does strict event grade filtering such as VFAINT mode on Chandra. For the instrumental background, one typically excludes events with energies near prominent instrumental lines, or at low energies ($E \lapprox 0.5$ keV) where the instrumental background turns upward. These can be effective at removing a good fraction of the unfocused background, but at the cost of discarding a considerable fraction (tens of percent) of the total events. Moreover, in observations of the faint emission around bright sources, VFAINT mode filtering could conceivably introduce a bias,  since it can exclude real events in regions where the count rate is high. 

Even with the standard reduction techniques, in most observations some unfocused background inevitably remains, and failure to explicitly account for this can lead to incorrect results. As an illustration we examine a recent 20 ks XMM-Newton EPIC observation of NGC 1961 (obs id 0723180101), a giant spiral galaxy surrounded by a hot gaseous halo (\citealt{Anderson2011}; \citealt{Bogdan2013}). The galaxy is placed at the aimpoint of the telescope for this observation. We reduce the data according to the procedure of Snowden and Kuntz (2013) to produce a PN image in the 0.4-1.25 keV energy band (a band which avoids the instrumental line at 1.5 keV and the uptick in the instrumental background below 0.4 keV).

The X-ray emission from the galaxy and its corona is visible to $\sim 3$ arcminutes, at which point it becomes indistinguishable from the background. The background is not flat however; it increases roughly from about 5 arcminutes out to the edge of the image. This is caused by the unfocused background. Dividing the counts per pixel by the exposure map (units of cm$^2$ s count photon$^{-1}$) implicitly assumes all the counts are distributed according to the exposure map, i.e. vignetted. Applying this vignetting correction to events associated with the unfocused background results in an overcorrection, causing the apparent increase in the background at large off-axis angle.

\begin{figure}
\begin{center}
\includegraphics[width=0.5\textwidth]{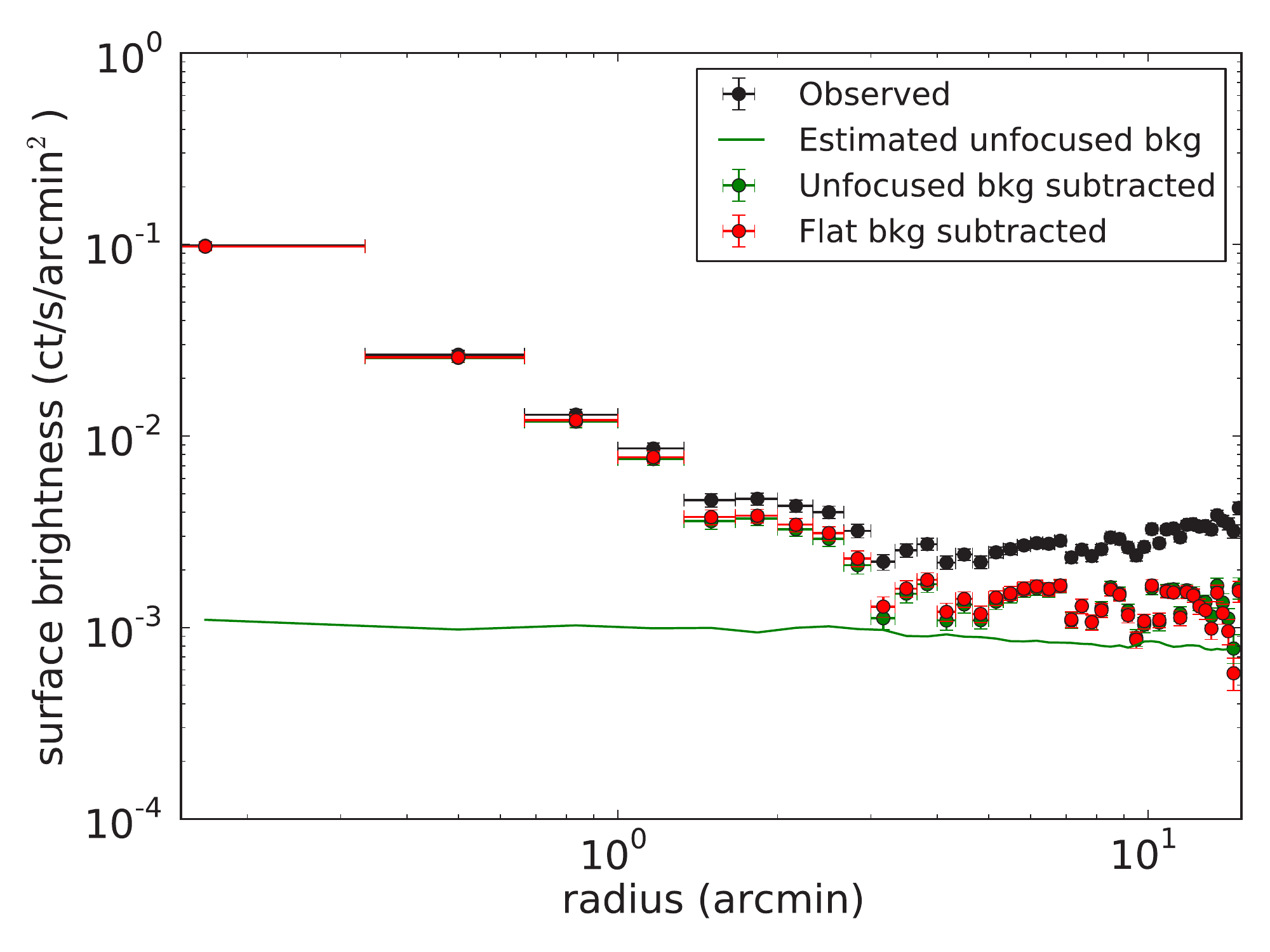}
\caption{XMM-Newton PN 0.4-1.25 keV radial surface brightness profile for a 20 ks observation of the giant spiral galaxy NGC 1961.  Black points are the vignetting-corrected observed data; note the overcorrection at $r>5$'. The green line is the estimated unfocused background (particle background + soft protons) estimated from the OFOV data + standard blank field templates. Subtracting the green line from the data before applying the vignetting correction yields a properly flat background (green points). The red points are the result if we assume the unfocused background is completely uniform, and they are not significantly different from the green points, suggesting this assumption is fairly good for this observation.    }
\end{center}
\end{figure}

If one naively divides by the exposure map without accounting for the unvignetted background, one would overestimate the background by as much as 50\% at 3 arcminutes. More likely, one could estimate the background based on, say, the 5'-7' annulus where the signal is fairly flat. This should theoretically produce the same background-subtracted profile as one obtains with the correction for the unvignetted background, at the cost of a less precise estimate of the level of the background due to the use of fewer photons. In this image, our band contains 538 photons in the 5'-7' annulus, so statistical uncertainties limit the precision of the estimation of the background to 4.3\%, even before considering any systematics. 

Since this is an XMM-Newton observation, we can use the \verb"pn_back" and \verb"proton" utilities to estimate the unfocused background in-field. We illustrate this component with the green curve in Figure 2. We can subtract it from the total emission (black) before the vignetting correction, and the resulting surface brightness profile (green points) no longer shows the uptick at large radii. Using the full 5'-14' region, after subtracting the unfocused background, the statistical uncertainty on the background is reduced to 1.6\%. 

In most situations, statistical uncertainty of 4\% for the background is perfectly adequate, but there are important exceptions. For example, in measurements of particularly faint emission (i.e. at \lapprox 10\% of the background), uncertainties in the background can be significant. Also, if the emission is particularly diffuse (extending more than 4', for example), there may be nowhere in the image where a flat in-field background can be measured. Finally, the particle and instrumental background are variable, and observations with higher unvignetted backgrounds will have correspondingly larger deviations from flatness when divided by the exposure map.  

Note that this unfocused background is nearly uniform.  If we perform the same analysis with the unfocused background modeled instead as a uniform distribution with the same average value, we obtain the red points in Figure 2, which are equal to the green points within the statistical errors. This simplifying assumption therefore seems fairly reasonable for this observation. As Figure 1 shows, the unfocused background is essentially uniform on the four Chandra ACIS-I chips, although it exhibits some structure on the ACIS-S chips. The shape is very different than the vignetting profile, however, so in the rest of this paper we will refer to the unfocused background as the ``unvignetted'' background. Since the unvignetted background is in general not known {\it a priori}, one must estimate it in-field. In the absence of OFOV data, this can be done by fitting to the observed surface brightness profile. One can then subtract it out, or include it in a model of the image. We discuss the latter option more in the next section.

\section{Image Modeling}

Instead of fitting and subtracting the unvignetted background, the more sophisticated approach towards accounting for the vignetted and unvignetted backgrounds is image modeling. In this study we construct a simple generic model for an X-ray image with an extended source. The model is constructed in the 0.5-2.0 keV energy band (which comprises the 'soft' and 'medium' bands as defined by the Chandra Source Catalog; \citealt{Evans2010}), and uses only spatial information (no spectral or temporal information, except for excluding events outside the energy band and excluding flaring events). Our model consists of three major components: the background, the source emission, and X-ray point sources. 

In this paper we use our model to study data taken with the Chandra ACIS instrument, so the emphasis will be on this instrument. It is fairly straightforward to implement this general model for XMM-Newton or for Suzaku, which may be done in future work. Below, we discuss each component of our model in turn, including details of applying each component to Chandra ACIS. 

\subsection{Background}
The model background has two components: the vignetted background and the unvignetted background. Each of these components is itself a mixture of a number of spectral components (as discussed in Section 2), but here we focus on their spatial distribution. The vignetted background (Cosmic + Galactic + Solar wind charge exchange X-rays) are focused by the telescope's optics and are therefore vignetted at large off-axis angles. The magnitude of this vignetting is energy-dependent, and we assume the vignetting profile is proportional to the exposure map in our energy band.  The shape of the exposure map has a weak dependence on the assumed spectrum of the incident photons; we assume a power-law distribution but we tried other simple models and found no significant differences\footnote{The {\it values} in the effective exposure map can vary significantly for different assumed spectra (by tens of percent), but we found that the {\it shape} (i.e. the vignetting profile) hardly changes at all.}. In general this is important to check for each energy band, since a difference in the vignetting profile for the background and the source photons could significantly bias the results of a spatial analysis. In practice, as long as the energy band is narrow enough that the instrument response is fairly uniform within the band, this should not be a major issue. Since the long-term goal is to extend this method to leverage spectral information as well, the expectation is that exposure maps will be constructed in many different energy bands across the energy range of interest, and it will be important to ensure that each energy band is sufficiently narrow. 

The unvignetted background was discussed in Section 2 above. It consists of cosmic rays, soft protons, and the instrumental background. In detail these backgrounds all have some spatial structure, but they are nearly flat when averaged over the 0.5-2.0 keV band (Figures 1 and 2).  In general we will treat this component as a uniform background, except when dealing with the ACIS-S chips where we use a heavily smoothed version of the image in Figure 1 as a template for the unvignetted background on these chips. 

\subsection{Source and Point Spread Function}
It is not necessarily required to parametrize the source emission, depending on the research objective. We develop two implementations of our image model: the {\it parametric} model, where we model the entire image and include a parameterization of the extended source emission, and the {\it nonparametric} model, where we model the background in the image but exclude the region around the source. The latter method therefore does not include a component for the source or the point spread function. It also has fewer data available for constructing the model, since the source region is masked, but once the model is constructed it can be subtracted from the full image, and the difference between the model and the data is the source emission and can be studied independently of any particular parameterization. 

For the parametric method, we opt to describe the source emission with a $\beta$-model \citep{Cavaliere1976}, which describes an isothermal, azimuthally symmetric gas distribution. In this model, the gas has an assumed density distribution

\begin{equation}\rho(r) \equiv \rho_0 \left[1 + \left(\frac{r}{r_0}\right)^2\right]^{-\frac{3}{2} \beta}\end{equation}

and the projected surface brightness profile of the X-ray emission as a function of projected radius $r$ is

\begin{equation}S_x(r) \equiv S_0  \left[1 + \left(\frac{r}{r_0}\right)^2\right]^{\frac{1}{2} - 3 \beta}\end{equation}

This profile is standard for describing the observed X-ray emission from gas in galaxy clusters and in hot halos around individual galaxies. It generally gives a good fit to the data and can be extended if necessary to account for radial temperature gradients, although for hot halos these gradients tend not to be very large. Another possible class of density profiles are the adiabatic profiles (\citealt{Maller2004}, \citealt{Fang2013}), which assume the gas is adiabatic instead of isothermal. This tends to produce a profile with a much flatter slope, which yields much lower surface brightness and is therefore difficult to constrain observationally. These profiles are employed less frequently than beta models, but for galactic hot halos they are still consistent with observations, and may be considered in future work. 

The source emission is convolved with the instrumental point spread function (psf), and for extended emission when we are trying to measure a surface brightness profile this can be an important effect. The Chandra psf is much smaller than the XMM-Newton psf or the Suzaku psf, subtending less than an arcsecond at the aimpoint, but it grows to several times this size at larger off-axis angles. In order to convolve our source emission model with the psf, we therefore need a functional form for the psf which is defined at every location on the detector. For this purpose, we specify a two-dimensional, azimuthally symmetric Gaussian at each point, so that the psf can be parametrized as a function of its standard deviation $\sigma$ at each point. We estimate $\sigma$ using the \verb"mkpsfmap" function, which estimates the radius corresponding to a given ecf (encircled counts fraction) at every location across an image\footnote{The function {\it mkpsfmap} does neglect aspect variations over the image, which can also broaden the effective psf, but these variations are uniform, random, and on the order of arcseconds, so they are not a major source of concern for this sort of analysis.}. We set the ecf to 0.393 so that the resulting radius is equal to $\sigma$. As with the exposure map, the psf depends on the spectral shape of the emission. We are generally interested in extended sources which are composed of hot gas and/or compact objects, but as long as the energy band is narrow the effect of the assumed spectral shape is minor\footnote{In fact, for our 0.5-2.0 keV energy band, and for the particular extended sources we study in this paper, we find that our results are actually the same within $1\sigma$ whether or not we convolve the image model with the psf at all.}.

In Appendix 1, we examine the nongaussianity of the psf, and we find that the Chandra psf in our waveband is very nearly Gaussian except at the aimpoint, where its small size makes the deviations from Gaussianity unimportant. 

We convolve our source emission model with the psf model to generate the total image model\footnote{In detail, we take the fast Fourier transform of the image and of the model for the spatially varying psf, multiply the two together, and then take the inverse Fourier transform. This is mathematically very similar to a convolution (identical in the limit of infinite spatial resolution), and is much quicker to compute.}. We also multiply the image model by the vignetting profile obtained from the exposure map. We do not convolve the psf model with the background, since the background is assumed to be uniform, and we do not convolve the psf model with the point sources described in the next subsection, since the point sources are computed with a more exact method, and so the approximation of Gaussianity is not needed. Therefore, the non-parametric method makes no use of the psf at all.

\subsection{Point sources}

In addition to the diffuse backgrounds, X-ray images are also contaminated by resolved point sources, and any image model should account for these sources as well. We take a two-pronged approach towards point sources in our image model. We attempt to mask the bright point sources where possible, and we study the completeness of our masking algorithm so that we can include a component in our image model to account for undetected point sources. 

\subsubsection{Masking Sources}

To detect bright point sources for masking, we run \verb"wavdetect" on the simulated images. We set the wavelet radii to 1, 2, 4, 8, and 16 pixels (with the image binning at 2, so one pixel is 0.984 arcseconds on a side), and the significance threshold to $10^{-6}$, corresponding to approximately one false positive per chip. The one non-standard setting we use with \verb"wavdetect" is to increase the size of the ellipse to mask ({\it ellsigma}) from $3\sigma$ to $8\sigma$ (the exact number here is not important, as long as it is large). This excludes nearly all of the emission from the point source, as well as a larger region around the point source than is typical. Our goal is to measure the background precisely, not to capture all of the background photons, so it is an acceptable tradeoff to lose some background photons in order to exclude as many of the photons from the bright point sources as possible.

During this process, we noticed that  \verb"wavdetect" was failing to detect some point sources at large off-axis angles (especially on the S2 chip). In order to study this effect further, we also tried an alternative method of detecting point sources. For every photon on the detector, we computed the 90\% encircled counts radius at the location of that event (using \verb"mkpsfmap" and assuming the same $\Gamma = 2$ powerlaw). We compared the number of counts within this radius to the estimated background (estimated either assuming a 100\% unvignetted fraction or an 100\% vignetted fraction; both methods give very similar results). If any circle has an excess of photons significant at $10^{-6}$ or stronger, we masked that circle.

This method is much more computationally intensive than \verb"wavdetect", since it has to perform aperture photometry tens of thousands of times (once per photon). It also has no ability to find the centroid of a point source; instead it just masks out any photons that could conceivably be associated with the source. While these constraints are not ideal for a robust multi-purpose point source detection algorithm, for our purposes we found this method to be complementary to \verb"wavdetect". In Appendix 2, we show a side-by-side comparison of point sources detected with the two methods in a single 100 ks observation of the Chandra Deep Field-South (CDF-S). In Appendix 3, we consider the effect of the choice of point source detection algorithm on our ability to recover input parameters. We find that both methods yield fairly similar results, but that the best results are obtained by using both methods on the image and masking sources with detections by either method.

\subsubsection{Accounting for Undetected Sources}

We account for undetected sources by simulating point sources in our image with a realistic N($>S$) distribution. We used the point source number counts function from the CDF-S 4 megasecond field \citep{Lehmer2012} for this distribution, which agrees well with previous estimates but has better sensitivity to sources with fluxes below $10^{-17}$ erg s$^{-1}$ cm$^{-2}$. 

Integrating their $dN/dS$ and plugging in the length of a given Chandra observation, we estimate the expected number of point sources in the image as a function of counts (assuming a $\Gamma = 2$ powerlaw spectrum to convert from flux to count rate). We generate a psf map using  \verb"mkpsfmap"  (using the 90\% ecf as the characteristic psf radius). For every integer number of counts in a given point source, we use the psf map to determine where in the image a point source with that number of counts could be detected above a given probability threshold (which we set to the value used by  \verb"wavdetect"  - $1\times10^{-6}$). We distribute the expected counts from undetected point sources over the area in which the point source would not be detected, and add together all the point sources expected in the image to get a map of the expected number of counts from undetected point sources at any given location in the field. An example of such a map is shown in Figure 3.

\begin{figure}
\begin{center}
\includegraphics[width=0.5\textwidth]{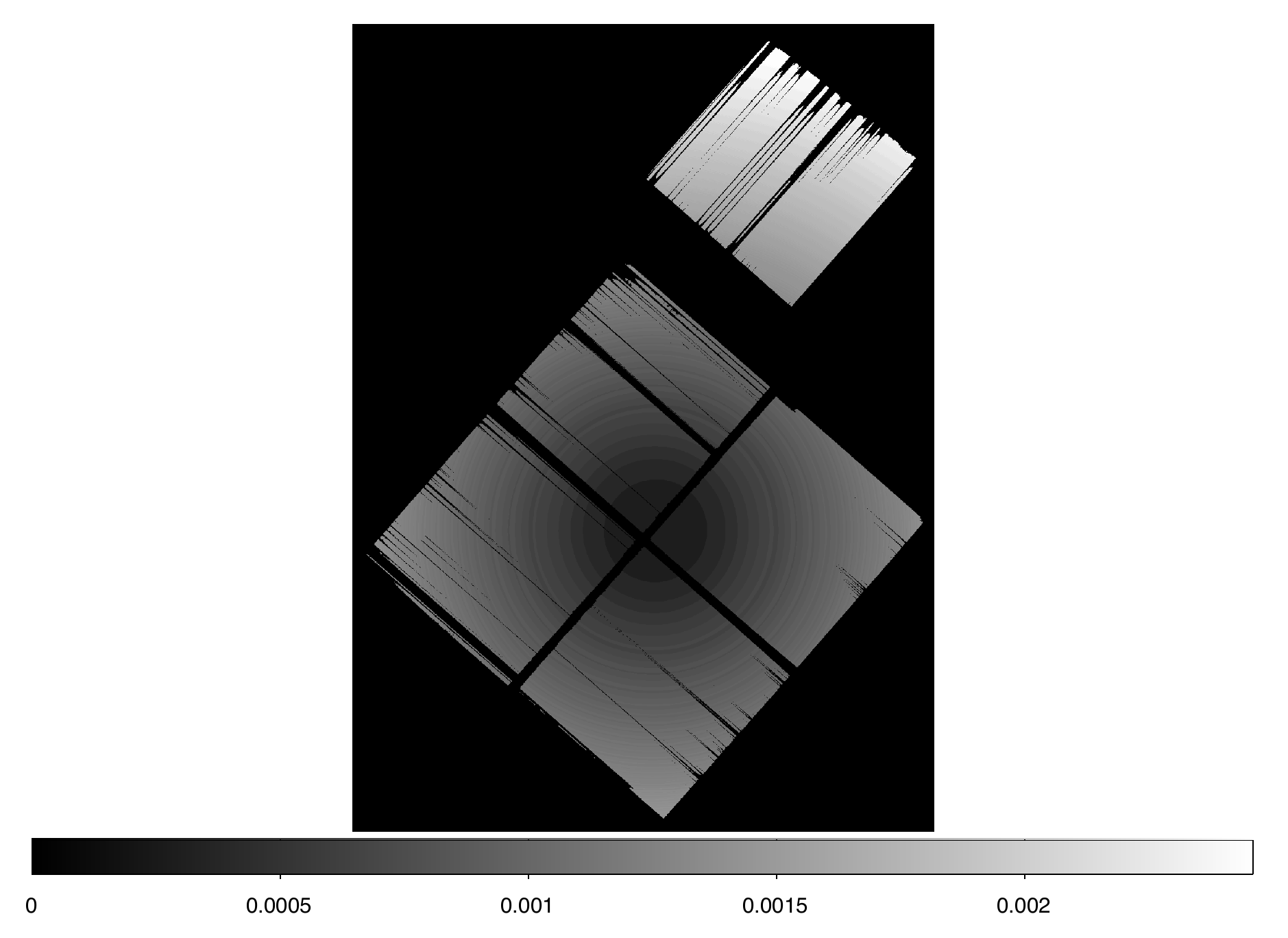}
\caption{Example map of the average number of counts per pixel corresponding to undetected point sources. This map is included as one component of our full image model, and is almost entirely determined by the length of the observation, the ACIS chip configuration, and the level of the background. The number of counts in this image corresponding to undetected point sources is 1136, and in any particular pixel the undetected point sources are a small correction (the total number of counts in the image is 52434). When binning the outermost pixels together, however, this component can be important. }
\end{center}
\end{figure} 

Note that this map is almost entirely determined by the length of the observation, the ACIS chip configuration, and the level of the background. In our experiments, we found that the dependence on the shape of the background is extremely weak:  the mean difference in the undetected flux per pixel between images with $f_{\text{unvig}} = 1$ and $f_{\text{unvig}} = 0$ is 0.3\%. This difference is so small because the two backgrounds only differ significantly at the largest off-axis angles, and at these angles most point sources are undetected everywhere so their flux is diluted across the entire image. Thus, we neglect the effect of the shape of the background on the distribution of undetected sources, so that this component is entirely determined by the length of the observation and the ACIS chip configuration, and therefore contains no free parameters. 

There are not enough point sources in a real observation to distribute their counts smoothly across every pixel, so we also explore the effect of binning the counts from regions with significant undetected point sources together (Appendix 2). This offers a modest improvement if performed correctly.

\section{Simulated Chandra Images}

Here we generate simulated Chandra ACIS images from our model, and study how well the model is able to recover the input parameters. 

\subsection{Generating Simulated Images}

Our fiducial model has five free parameters - two normalizations for the vignetted and unvignetted backgrounds, and the three parameters that define our source emission ($\beta$, $r_0$, and $S_0$).  We also need to specify shapes for the vignetted and unvignetted backgrounds, although these are completely determined by the details of the observation and contain no free parameters. In practice, we also fix the average flux per pixel to the measured value in a given observation, which removes one of the free parameters from the normalizations of the two backgrounds (since their sum is fixed). We therefore instead parameterize the backgrounds in terms of $f_{\text{unvig}}$, the fraction of the diffuse background counts that we attribute to the unvignetted component. In addition to these model components, we also add random point sources to simulated images as discussed below.

The vignetted background follows the shape of the exposure map generated with \verb"mkexpmap" (either in units of cm$^2$ count photon$^{-1}$ or cm$^2$ s count photon$^{-1}$, although we use the latter). For a given observation, \verb"mkexpmap" computes the effective exposure time and/or area across the image, so we can use its output as a vignetting profile for the observation. The inputs are an aspect histogram and an instrument map file. The aspect histogram encodes the movement of the telescope during the observation, and the instrument map encodes the quantum efficiency and detector response as a function of energy for the specified source spectrum (see the CXC helpfile or \citet{Davis2001} for more details). 

We used the  \verb"fluximage" script to create an exposure map, with the non-standard setting of  \verb"expmapthresh = 20%" instead of the more common value of 2\%. This zeros out every pixel with effective exposure time less than 20\% of the peak value, which in practice eliminates some pixels near the edges of chips where the calibration is more uncertain. The vignetting across the ACIS chips is typically fairly modest, so this has little effect except near the edges of the chips. However, it has the advantage of removing from our analysis the chip gaps, where the exposure map is the most uncertain. We can then define the shape of the unvignetted background to be uniform across the image wherever the exposure map is non-zero.

We also add point sources to the image in order to evaluate any systematic effects they may introduce to the analysis. While we rely on \verb"mkpsfmap" for modeling the psf, this is necessarily an approximation, so when constructing simulated images we attempt to simulate point sources as accurately as possible. The shape of the Chandra psf varies as a function of location on the detector and of energy, so it is not trivial to generate a simulated point source in detail, but this process has been automated with the Chandra Ray Tracer (ChaRT; \citealt{Carter2003}) and the Model of AXAF Response to X-rays (MARX; http://space.mit.edu/cxc/marx/) software. We used ChaRT to construct a library of psf shapes across the ACIS field of view. We generated simulated psfs in polar coordinates, recording a psf at every arcminute of off-axis angle (also sampling every half arcminute near the aimpoint) and every  $30^{\circ}$ of position angle. We used a $\Gamma = 2$ powerlaw for the spectral shape of the point sources over the 0.5-2.0 keV band. We then passed the simulated psfs to MARX, and traced $10^4$ photons through the telescope optics in order to construct an events file for a point source at each location. We repeated this process for both the ACIS-I and ACIS-S configuration, as described in the MARX help files, in order to be able to simulate point sources on the ACIS-S chips in an ACIS-I observation, or vice versa.

Armed with this library of events files, we can simulate a point source at any location in an observation. We find the nearest point source events file from our library, center it on the location we want to simulate a point source, and draw events from the events file at random in order to build up the desired count rate. We use the point source number counts function estimated from \citet{Lehmer2012} to estimate the number and count rate of point sources to generate for a given image.

We can now simulate an X-ray image. The background is generated by multiplying the vignetted background flux by the vignetting profile specified by the exposure map, multiplying the unvignetted flux by the area of the image specified by the exposure map, and combining the two. For simulated images which include the ACIS-S chips, we use a smoothed version of the image in Figure 1 as a template for the shape of the unvignetted background across the ACIS-S chips. We then add an extended source component to the image, convolved with the spatially varying Gaussian Chandra psf. Finally, we add point sources randomly to the image with fluxes and shapes determined as described above.

We can derive a simple prediction for $f_{\text{unvig}}$ from estimates of the cosmic X-ray background and the average instrumental background. From \citet{Markevitch2003}, the average instrumental flux rate in the 1.0-2.0 keV band is about $1.1\times10^{-15}$ erg s$^{-1}$ cm$^{-2}$ arcmin$^{-2}$ (there is additional flux below 1 keV, but it is highly variable). Comparing this to the average cosmic X-ray background of $2.1\times10^{-15}$ erg s$^{-1}$ cm$^{-2}$ arcmin$^{-2}$ \citep{Bauer2004}, the first-order estimate for the unvignetted fraction in a random field is at least 0.34, plus an additional component based on the $<1.0$ keV instrumental flux at that time. However, the unvignetted background and the SWCX component of the vignetted background are both variable, so significant variations in $f_{\text{uvig}}$ are expected. The sensitivity of the flaring correction can also affect $f_{\text{uvig}}$, since a higher fraction of the events during flares are from the non-X-ray background and therefore unvignetted. Additionally, the backside-illuminated chips are more sensitive to low-energy instrumental photons, so observations in the ACIS-S configuration might be expected to have a higher value of $f_{\text{unvig}}$. Finally, the use of VFAINT mode filtering will generally reduce $f_{\text{unvig}}$. 

In Figure 4 we illustrate three example simulated images. Two images have been produced with the aimpoint on the ACIS-I3 chip, and one with the aimpoint on the ACIS-S3 chip. We also add an extended source to each image. For one of the simulated images in the ACIS-I configuration, we place the extended source far off-axis, on the I0 chip. For the other two simulated images, the extended source is near the aimpoint. Each image has the same parameters:  $f_{\text{uvig}} = 0.4$, $\beta = 0.5$, $r_0 = 10$ arcsec, and $N = 2000$\footnote{$N$ is defined as the number of counts in the extended source, integrated out to a radius of 400 pixels.}. We apply the same average background flux to each image of $0.035$ ct pix$^{-1}$, corresponding to the typical 0.5-2.0 keV background in an observation of $\sim 100$ ks. 

\begin{figure*}
     \begin{center}
            \includegraphics[width=0.9\textwidth]{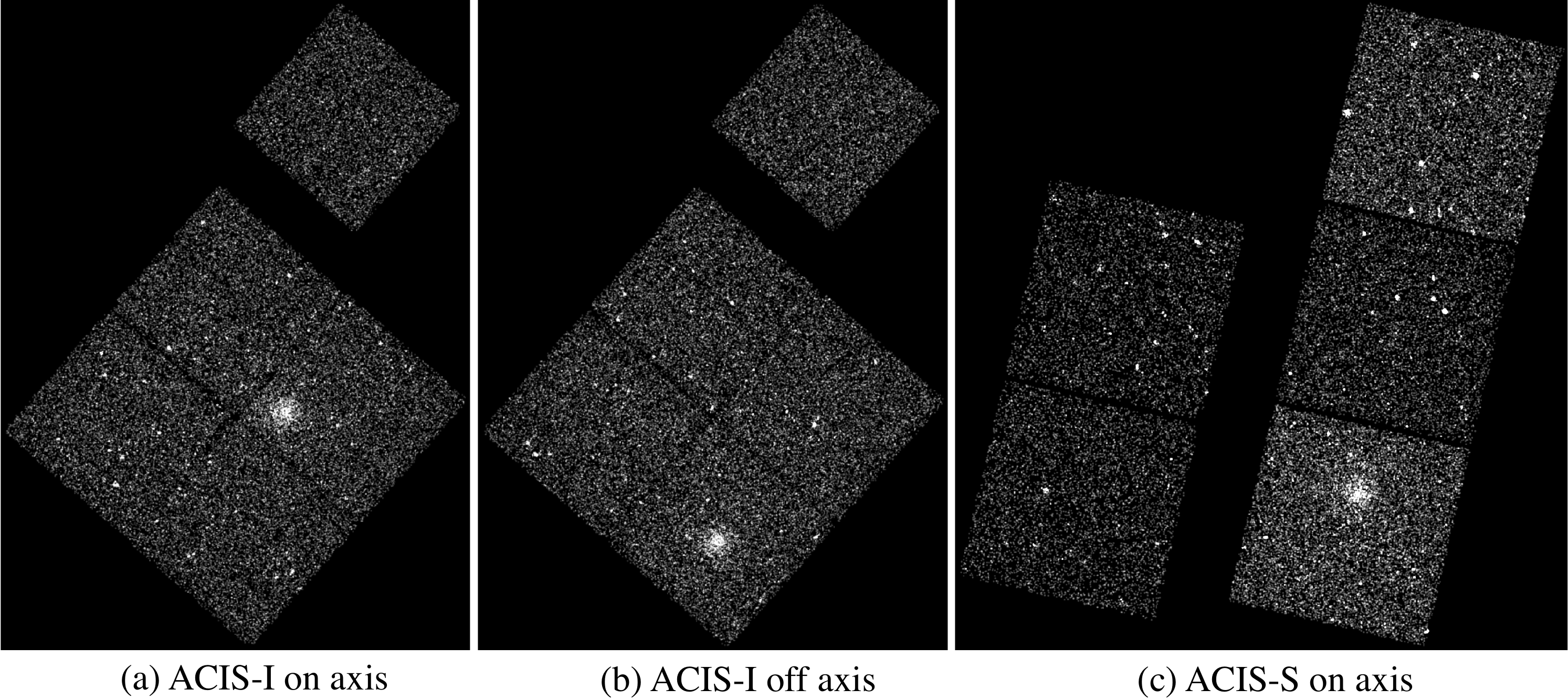}
    \caption{\small Simulated images with mock point sources added at random, and an extended source with parameters as described in section 4.1. Images (a) and (b) have configurations with the aimpoint on ACIS-I3, image (c) uses a configuration with the aimpoint on ACIS-S3. The extended source is nearly on-axis for images (a) and (c), and the source is far off-axis in image (b). All three images have been smoothed with a 3-pixel Gaussian kernel. }
         \end{center}
\end{figure*}

\subsection{Likelihood Function}

We need a likelihood function to relate this (simulated) data to image models. In Appendix 3, we test a variety of likelihood functions and settle on a likelihood function which evaluates the likelihood at every pixel, with some binning at large off-axis angles. The full likelihood function is:

\begin{equation}L \equiv \prod_{\text{pixels}} p(c_{\text{pix}} | m_{\text{pix}}) + p(c_{\text{bin}} | m_{\text{bin}}) \end{equation}

where every pixel is evaluated separately, except for the 10\% of pixels for which the psf is largest. Those 10\% of pixels are combined into one bin. Then $p(c|m)$ is the Poisson probability of obtaining the observed number of counts $c$ in a pixel/bin, given a model prediction $m$ for that pixel/bin. 

\subsection{Recovering Input Parameters - Nonparametric Method}

We first examine the data nonparametrically. We mask out point sources using a combination of \verb"wavdetect" and our manual method, and we exclude everything out to a radius of 5' around the extended source in our image. The goal is to use the remaining data to estimate the vignetted and unvignetted backgrounds, which can then be added to the estimated undetected point source component to produce a full model background image. We can then subtract this background model from the real image (with point sources again masked, but the extended source unmasked) in order to obtain an estimate of the background-subtracted extended emission. 

We use the likelihood function from section 4.2 to determine the background model. Using maximum likelihood, we find the best-fit value of $f_{\text{unvig}}$ for each image, and then generate a model background for the full image, consisting of vignetted and unvignetted backgrounds, and the parameter-free component accounting for undetected point sources at large off-axis angles. 

In Figure 5 we show the results of this analysis. We measure the surface brightness profiles around each of our three simulated images, divide by the exposure map, and then subtract the estimated composite background. We have fit an error-weighted smoothing spline to the profile data in order to minimize the effects of the binning and to show the magnitude of the negative points at large radius. For comparison, the true source emission is shown by the green line, and clearly the non-parametric method can recover the emission very well. 

\begin{figure*}
     \begin{center}
            \includegraphics[width=1.0\textwidth]{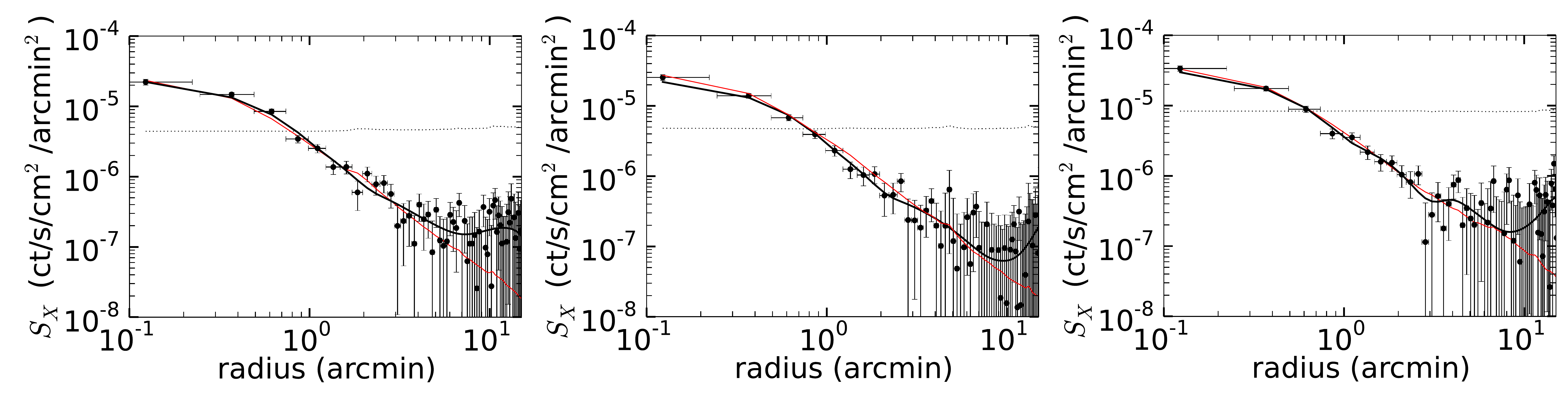}
    \caption{\small Background-subtracted surface brightness profiles of the extended sources in Figure 4. The background was estimated non-parametrically (section 4.3), and includes vignetted and unvignetted components; it is indicated with the dotted black line. The red curve shows the true surface brightness profile, i.e. the model profile used to generate the simulated image. Accounting for the shape of the background is necessary to match the model once the flux falls significantly below the background. The error bars account (in quadrature) for both uncertainties in the number of source counts (Poisson errors) and in the level of the background (based on the range of acceptable values for $f_{\text{unvig}}$), and are displayed at 1$\sigma$. A smoothing spline (black line) has been fit to the recovered profile data in order to minimize the effects of the binning and to show the magnitude of the negative points at large radius.}
    \end{center}
\end{figure*}

The simulated images in the ACIS-I configuration return values of $0.38_{-0.02}^{+0.01}$ and $0.40 \pm 0.02$ for $f_{\text{unvig}}$ in the off-axis and on-axis sources respectively. For the simulated image in the ACIS-S configuration, however, the non-parametric method returned an incorrect value of $0.17\pm0.02$ for $f_{\text{unvig}}$ (recall the true value in these simulated images is 0.40). This could be due to the bright regions of the backside-illuminated (BI) S1 chip being excluded due to point sources, or to missed point sources on the frontside-illuminated (FI) chips, either of which would bias $f_{\text{unvig}}$ downwards; however, when we perform a parametric analysis of this simulated image in the next section, the recovered $f_{\text{unvig}}$ is much closer to the true value. This suggests that the issue is related to the additional exclusion region imposed by the non-parametric analysis. Almost the entire BI S3 chip is excluded by our 5' source exclusion region, and most of the BI S1 chip is excluded due to point source emission, so the dynamic range over which to discriminate the vignetted and unvignetted background is more limited, and subtler biases related to point sources can play a larger role here. Fortunately, since the difference between the unvignetted and vignetted backgrounds is smaller in the field of view for this configuration, the value of $f_{\text{unvig}}$ is less important. On the other hand, the uncertainty in the background subtraction is poorer, as can be seen by the higher average value for the noise at large radii for the ACIS-S configuration in Figure 5. 

We compare the results above to a standard analysis which does not account for the shape of the background. We divide the images (Figure 4) by the exposure map, yielding curves like those in Figure 2 which turn upwards at large radii due to the unvignetted background. We select an annulus where the background is at a minimum and appears closest to flat, and estimate the background in-field within this annulus. The surface brightness of this annulus is assumed to be the surface brightness of the background across the image. We then subtract this value from the image, yielding the profiles in Figure 6.

\begin{figure*}
     \begin{center}
            \includegraphics[width=1.0\textwidth]{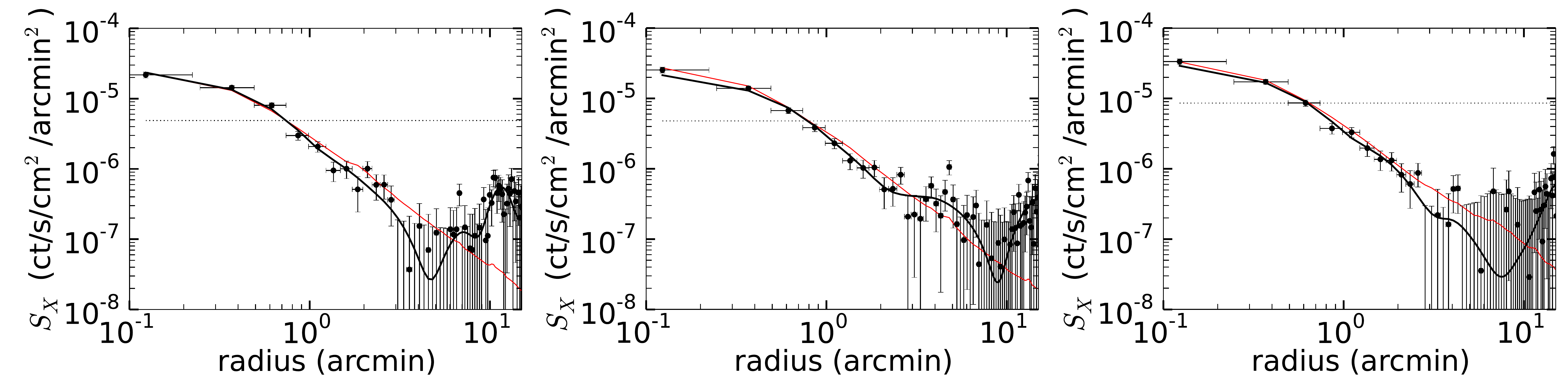}
    \caption{\small Background-subtracted surface brightness profiles of the extended sources in Figure 4. Here the background was estimated using a standard in-field analysis which does not account for the unvignetted background; it is indicated with the dotted black line. The red curve shows the true surface brightness profile, i.e. the model profile used to generate the simulated image. This causes systematic failures in matching the true surface brightness profile once the signal falls roughly below the background. The error bars account (in quadrature) for both uncertainties in the number of source counts (Poisson errors) and in the level of the background (Poisson errors from the number of counts within the in-field background annulus), and are displayed at 1$\sigma$.  A smoothing spline (black line) has been fit to the recovered profile data in order to minimize the effects of the binning and to show the magnitude of the negative points at large radius. Comparing to Figure 5, the standard in-field analysis performs noticeably worse if one tries to follow the signal below the background. }
        \end{center}
\end{figure*}

The differences between the source and the recovered profile are negligible within an arcminute, where the source emission is well above the background. Once the source becomes comparable to, or below, the background, these differences are more important. For the ACIS-I on-axis source, the source emission is lost at 3', while there is still evidence of signal at 5' when we model the background. The off-axis source yields a false upturn in the source emission at 4-5', which is not present when we model the background. In the ACIS-S configuration, the background is over-estimated, so the recovered signal is systematically below the true signal at radii beyond 1'.

We can quantify these differences by calculating the $\chi^2$ goodness of fit parameter for each simulation. We calculate $\chi^2$ out to the radius at which the source is entirely lost to the background for both the modeled and the flat background analyses. This is 6' for the ACIS-I sources and 4.5' for the ACIS-S source. For the ACIS-I on-axis source, we find $\chi^2 / $d.o.f. =  26.4/24 and 40.7/24 for the modeled and flat backgrounds, respectively. For the ACIS-I off-axis source, $\chi^2 / $d.o.f. =  18.4/24 and 35.7/24 for the modeled and flat backgrounds, respectively. For the ACIS-S source, $\chi^2 / $d.o.f. =  16.9/17 and 21.6/17 for the modeled and flat backgrounds, respectively. In each case, we find an acceptable match between the data and the source model if we model the background, and a much less acceptable match if the background is allowed to be flat (although the difference between the two is much smaller for the simulation in the ACIS-S configuration).

The major drawback to the non-parametric method is that it requires the source emission to be fairly localized within the image so that it can be masked out in order to fit the background. This is not always possible, and motivates the use of a parametrized model for the source in such cases, so that the background can be fit from the full image without having to mask the region around the source.

\subsection{Recovering Input Parameters - Parametric Method}

Here we perform a similar analysis, but without masking the region around the extended source. Now the model has four free parameters instead of just one, so running the maximum likelihood analysis becomes much more cumbersome. We found that a fixed grid search was too inefficient for exploring this space, so we opted for Markov Chain Monte Carlo (MCMC) instead. An MCMC search is a Bayesian technique designed to efficiently sample likelihoods in multidimensional space. For long enough chains, the distribution of samples in multidimensional space is supposed to approximate the posterior probability distribution for the fit parameters. Obviously, since it is a Bayesian technique, MCMC requires specification of priors for each fit parameter as well. 

In this analysis, we use the open-source Python package \verb"emcee" \citep{Foreman-Mackey2012}, which implements an affine-invariant MCMC method (Goodman and Weare 2010). This implementation of MCMC uses a user-specified number of independent ``walkers'' which separately explore the parameter space. The prior sets a uniform region, from within which each walker is randomly assigned an initial position. But as long as the chain is run for a sufficiently long time, and the likelihood function is well-behaved, the walkers can explore the entire space and the result is nearly completely insensitive to the choice of prior. 

We used uniform broad priors on $\beta$, $r_0$, $N$, and $f_{\text{unvig}}$, and used the same priors for each simulation. We also ran MCMC chains with different priors, including narrower priors around incorrect values, and recovered the same posterior probability distributions. This allows us to verify that our choice of priors has no effect on the posterior probability distribution. We use 50 walkers and compute a chain of 1100 elements for each walker, discarding the first 1000 and keeping the final 100 for analysis.

In Figures 7-9 we show the posterior probability distributions for these four parameters, for each simulation. The full image modeling routine is successfully able to recover the input parameters, even for the case with the source on the ACIS-S chips. Interestingly, the full image model performs worse when the source is located on the aimpoint for an ACIS-I configuration (Figure 7) as compared to placing the source off-axis (Figure 8). This is probably due to the degeneracy between the vignetting profile and the surface brightness profile of the source, which makes it difficult to disentangle the shape of the source surface brightness profile.

For comparison, we also generate image models with $f_{\text{unvig}} \equiv 0$, which simulates the effect of ignoring the unvignetted background and simply dividing the cleaned image by the exposure map. For the ACIS-I configurations, models with $f_{\text{unvig}} = 0$ perform significantly worse than models where $f_{\text{unvig}}$ is fit along with the source parameters. The difference is much smaller for the ACIS-S configuration (Figure 9), because our source is largely confined to the BI chip where the difference between the vignetted and unvignetted backgrounds is small. A larger source would probably present more difficulty for a model with $f_{\text{unvig}} \equiv 0$. 

There is a discrepancy between the recovered value of $N$ and the input value for the simulated image with the ACIS-S configuration, significant at about the $1\sigma$ level. Since we present three simulated images, a $1\sigma$ discrepancy in one image is not very statistically meaningful, and in our experience we have noticed no systematic discrepancy in the recovered values of $N$ with the ACIS-S configuration.

\begin{figure*}
     \begin{center}
            \includegraphics[width=1.0\textwidth]{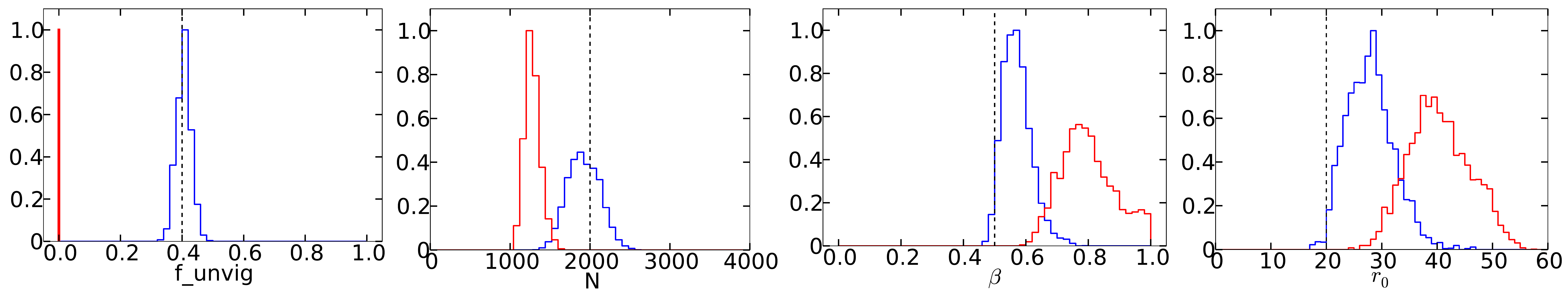}
    \caption{\small Posterior probability distribution functions (pdfs) for the image parameters corresponding to the ACIS-I on-axis simulated image in Figure 4a. The dashed lines are the true values of the parameter. The blue lines are the pdfs if we include $f_{\text{unvig}}$ in our fit; the red lines illustrate the effect of neglecting $f_{\text{unvig}}$ and setting it to zero, as in a standard spatial analysis.  Our method, which accounts for the unvignetted background, is strongly favored. }
    \end{center}
\end{figure*}

\begin{figure*}
     \begin{center}
            \includegraphics[width=1.0\textwidth]{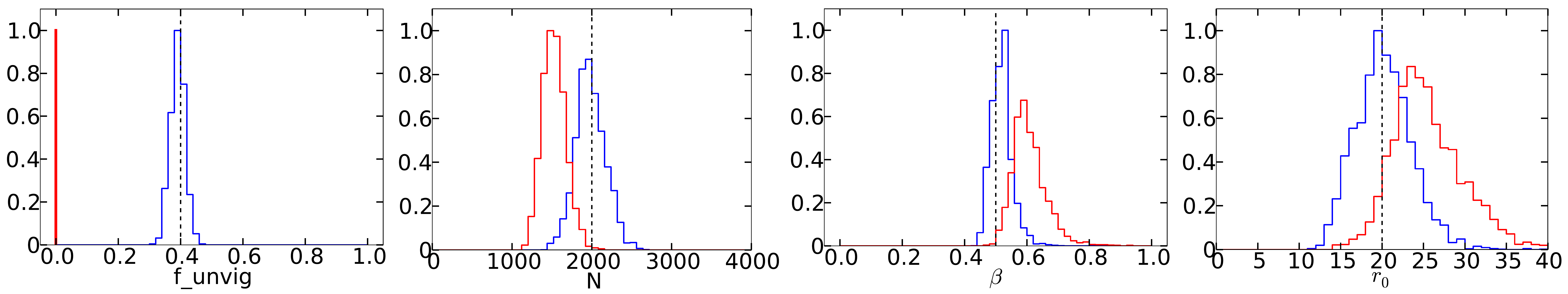}
    \caption{\small Posterior probability distribution functions (pdfs) for the image parameters corresponding to the ACIS-I off-axis simulated image in Figure 4b. The dashed lines are the true values of the parameter. The blue lines are the pdfs if we include $f_{\text{unvig}}$ in our fit; the red lines illustrate the effect of neglecting $f_{\text{unvig}}$ and setting it to zero, as in a standard spatial analysis.  Our method, which accounts for the unvignetted background, is strongly favored. }
    \end{center}
\end{figure*}

\begin{figure*}
     \begin{center}
            \includegraphics[width=1.0\textwidth]{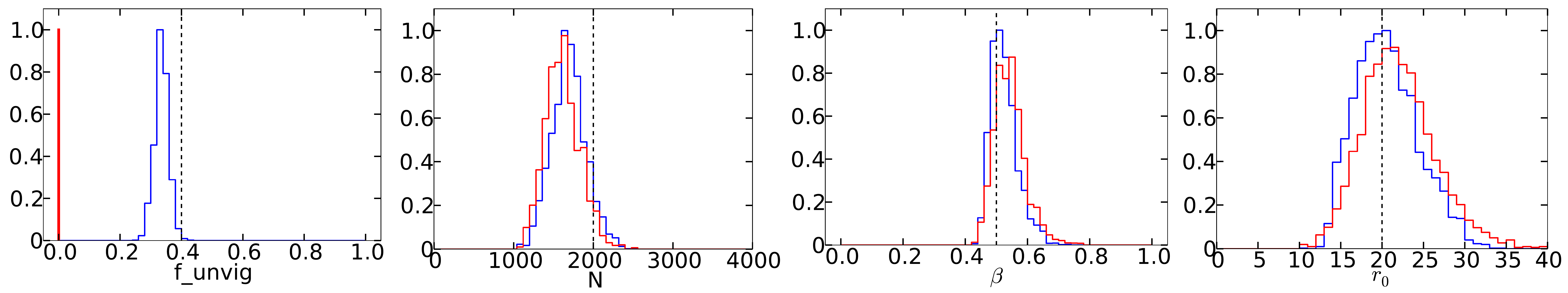}
    \caption{\small Posterior probability distribution functions (pdfs) for the image parameters corresponding to the ACIS-S simulated image in Figure 4c. The dashed lines are the true values of the parameter. The blue lines are the pdfs if we include $f_{\text{unvig}}$ in our fit; the red lines illustrate the effect of neglecting $f_{\text{unvig}}$ and setting it to zero, as in a standard spatial analysis.  Our method, which accounts for the unvignetted background, is slightly favored, but the difference is not very significant. For a source of larger radial extent, we expect the difference would be larger.  }
    \end{center}
\end{figure*}

\subsection{Summary of Simulations}

In this section we examined the behavior of both parametric and non-parametric image analysis methods, for simulated images of faint extended emission in three different Chandra ACIS configurations (two on ACIS-I, one on ACIS-S). The non-parametric method successfully recovered $f_{\text{unvig}}$ for both simulated images with the ACIS-I configuration, but it underestimated $f_{\text{unvig}}$ significantly for the simulated image with the ACIS-S configuration. The effect of $f_{\text{unvig}}$ does not seem very important for the simulated image in the ACIS-S configuration. However, in all three images, the non-parametric method provides a statistically acceptable match to the simulated source out to the maximum radius at which the source is detectable. This radius is smaller for the ACIS-S simulated image (4.5', compared to 6' for the ACIS-I simulated image).

For the parametric method, we specified a likelihood function for comparing an image model to data, and we showed that this likelihood function allows us to recover the parameters used to generate the simulated image. Parameters are recovered within the $1\sigma$ uncertainties in most cases, and these $1\sigma$ uncertainties are typically 10-20\% of the size of the parameter. The parametric method is better able to recover $f_{\text{unvig}}$ for the simulated image in the ACIS-S configuration, although this parameter has little effect on the other parameters for this configuration. For the ACIS-I configurations, on the other hand, mis-specifying $f_{\text{unvig}}$ causes very significant biases in the recovered values of the other image parameters. Fortunately, $f_{\text{unvig}}$ is recovered accurately in these images as well.

\section{Examining Chandra Data: NGC 720}

We have seen that this method can recover an extended surface brightness profile robustly in realistic simulations, for both ACIS-S and ACIS-I configurations. In this section we apply our analysis to real data. 

We examine the slightly larger than $L*$ elliptical galaxy NGC 720 (d = 25 Mpc, $M_K = -24.6$) . This galaxy was discussed briefly in Section 1 as an example of an object for which spectral fitting has produced inconclusive results. \citet{Humphrey2006} (hereafter H06) inferred a hot halo mass of $\sim 1\times10^{11} M_{\odot}$ around this galaxy, while \citet{Humphrey2011} (hereafter H11) inferred a hot halo mass about three times larger. It is notable that such different results were obtained for the same galaxy. The earlier paper is based on Chandra obs id 492, a 40-ks ACIS-S observation of the galaxy, of which only 17 ks were used due to flaring. The later paper uses different observations:  four other Chandra pointings with ACIS-S (obs id 7062, 7372, 8448, 8449) adding to 100.5 ks, and a 177 ks Suzaku pointing (obs id 80009010). According to H11, if they exclude the Suzaku data, the inferred hot halo mass decreases by $50\%$, which suggests that much of the discrepancy is driven by the Suzaku data (they emphasize XIS1 in particular). However, at smaller radii the discrepancy between Chandra and Suzaku essentially disappears, while the discrepancy between the two papers remains. 

We investigate this discrepancy in more detail by examining the radial density profiles for the hot gas, as derived in both H06 and H11. In Figure 3 of H06, they present their inferred deprojected gas density profile, extending out to 90 kpc. We read in this profile from their figures, including their $1\sigma$ uncertainties, and used \verb"emcee" to fit $\beta$ models to the data. The best-fit profile had $\beta=0.45$, $n_0 =  0.04$ cm$^{-3}$, $r_0 = 1.1$ kpc, yielding a hot gas mass within 300 kpc of $1.5\times10^{11} M_{\odot}$; this figure accords with their enclosed gas mass as displayed in their Figure 6. We retained the full range of fits from the MCMC chain in order to estimate the uncertainties around the best-fit profile as well. 

H11 provides a projected density profile, although their definition of ``projected density'' is fairly difficult to deproject and analyze. We instead are able to obtain a density profile from their Figure 6, which shows the cumulative enclosed gas mass as a function of radius, extending from about 30 kpc to 400 kpc. The cumulative enclosed mass profile in this range is very well fit with a power-law with slope $1.76\pm0.02$, corresponding to a $\beta$-model with $\beta = 0.41\pm0.01$. The core is not resolved, so we leave this as a free parameter. The normalization can be obtained from the enclosed mass in H11 Figure 6, for any given value of $r_0$. Unfortunately, the profile in their Figure 6 is just their best-fit profile, and the uncertainties around it need to be estimated as well. We estimate the uncertainties in this profile by scaling the uncertainties from H06 by the size of the error bars on the points in the projected density profile in H11. As H11 point out, the errors are much smaller for the density profile in their newer analysis, partially because they do not deproject the profile, but they are also subject to considerably larger systematics. 

Next, we analyze this galaxy with our methods. NGC 720 has six different Chandra observations - five in the ACIS-S configuration (obs ids 492, 7062, 7372, 8448, and 8449) with the galaxy on the S3 chip like our simulated images, and one (obs id 11868) in the ACIS-I configuration with the galaxy at a similar off-axis position to our ACIS-I off-axis simulated image. We reduced these six observations in the standard way, using  \verb"chandra_repro" to reprocess, and then  \verb"fluximage" to generate the images and exposure maps. As above, we ran \verb"fluximage" with the parameter \verb"expmapthresh" set to 20\% and we generated images in the 0.5-2.0 keV band. We run both  \verb"wavdetect" and our manual point source detection algorithms on each image, and mask out point sources detected with either method. 

\subsection{Analysis with the non-parametric method}

NGC 720 is somewhat larger and brighter than our simulated images, and subtends nearly an entire chip. This makes it more difficult to perform the non-parametric analysis when NGC 720 is placed on the S3 chip. We therefore only examine observation 11868, where the galaxy is placed off-axis on the I1 chip. We mask out a circle of radius 7' around the galaxy. This is a larger radius than in our simulations but a larger radius is necessary because of NGC 720's brightness and extent. We find $f_{\text{unvig}} = 0.05_{-0.03}^{+0.04}$, so this observation has nearly no unvignetted background. This observation has no significant flaring, and ACIS-I is less sensitive to the softer energies where the instrumental background is higher, which may explain the low value of $f_{\text{unvig}}$ in this observation. The background-subtracted surface brightness profile, and the best-fit level of the background, are shown in Figure 10a.

For comparison we also show a surface brightness profile estimated using a uniform vignetted background estimated in-field. Due to the size of the excluded region, there is no annulus available in-field around the galaxy which can be readily used for the background, so we instead select an in-field background from a roughly conjugate region on the I2 chip, in a similar method to \citealt{Anderson2011}.  The surface brightness profile measured in this way is shown in Figure 10b. The two profiles are nearly identical, which is not surprising given the low value of $f_{\text{unvig}}$ in this observation (note that if $f_{\text{unvig}}$ = 0, the non-parametric method reduces to a standard in-field background subtraction). Emission is detected securely out to at least 4 arcminutes (30 kpc), and possibly up to 6 arcminutes (45 kpc), depending on the size of the spatial bins. Note that these radii are much smaller than the outermost annulus at 11' in H06, and are much closer to the 6' outermost annulus in H11. 

\begin{figure*}
     \begin{center}
        \subfigure{%
           \includegraphics[width=0.5\textwidth]{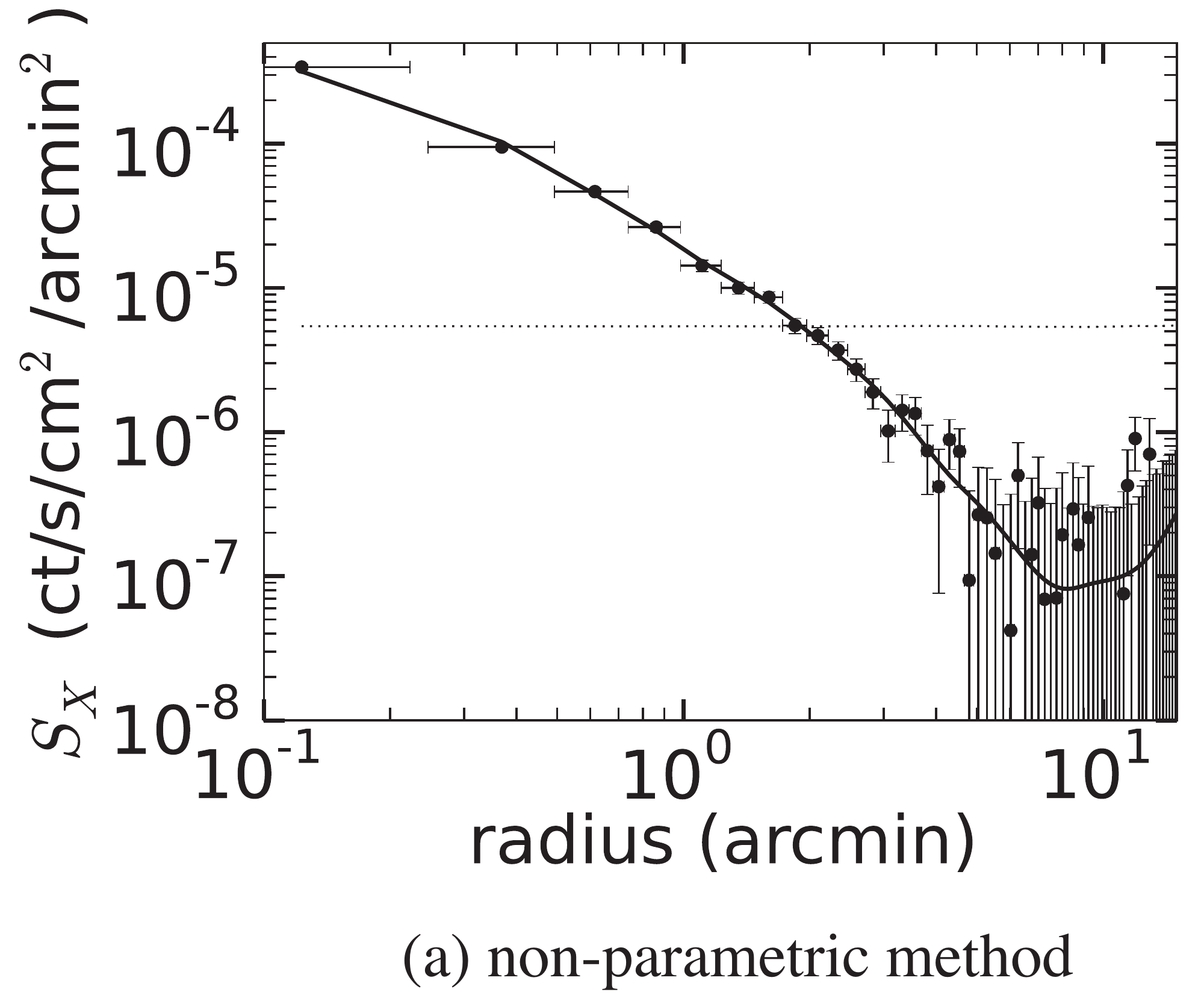}
        }%
        \subfigure{%
            \includegraphics[width=0.5\textwidth]{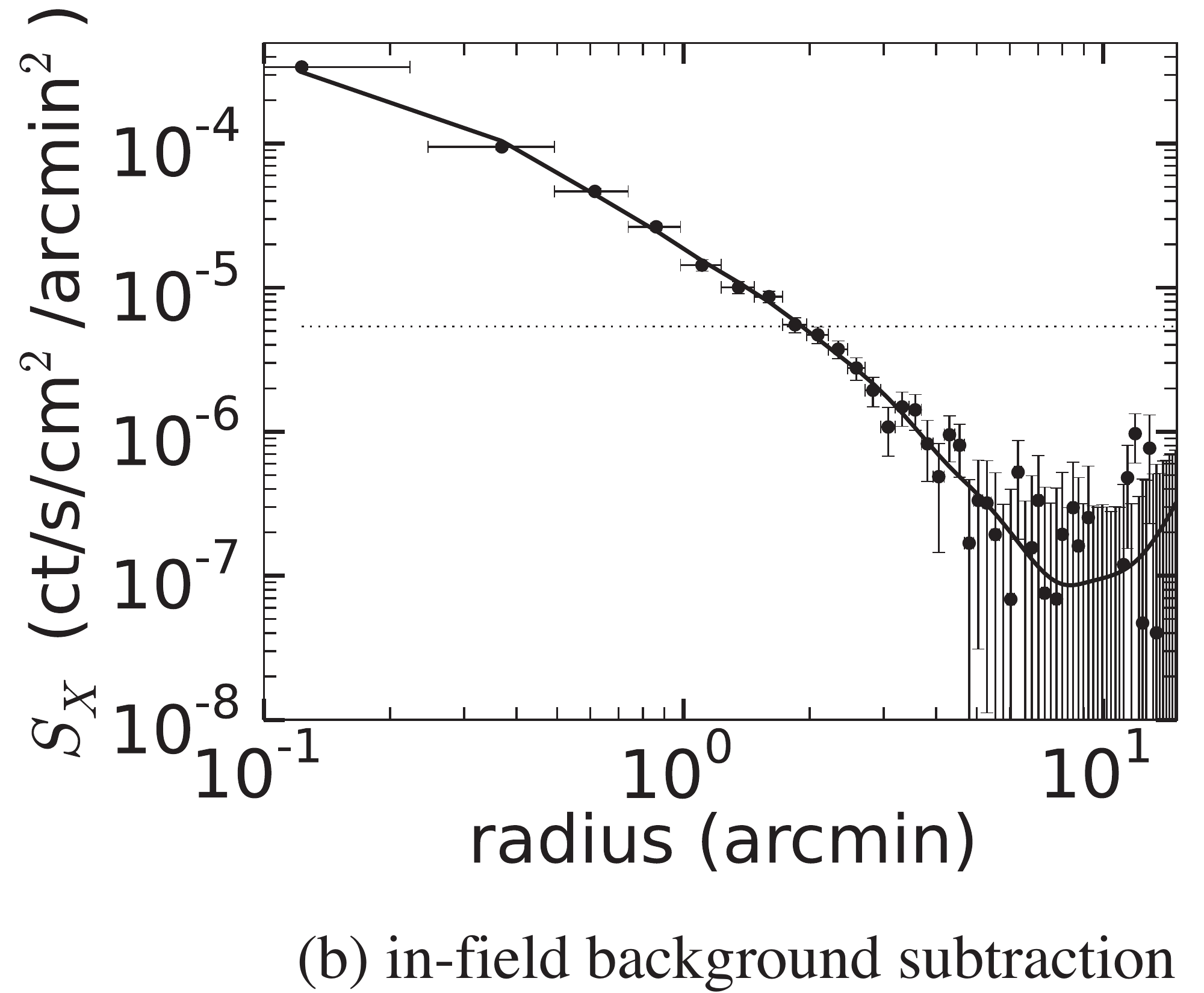}
        }%
    \caption{\small Background-subtracted surface brightness profiles of the emission around NGC 720, based on the off-axis observation in the ACIS-I configuration  (obs id 11868) In (a), the background was estimated using the non-parametric method in section 4.3, and has  $f_{\text{unvig}} = 0.05_{-0.03}^{+0.04}$. In (b), the background was assumed to be entirely vignetted and was determined using an in-field annulus. For both plots, the error bars account (in quadrature) for both uncertainties in the number of source counts (Poisson errors) and in the background, and are displayed at 1$\sigma$. A smoothing spline has been fit to the profile data in order to minimize the effects of the binning and to show the magnitude of the negative points at large radius. The two plots are essentially equivalent, since  $f_{\text{unvig}}$ is so close to zero that it has little effect on this observation. }
        \end{center}
\end{figure*}

\subsection{Analysis with the parametric method}

With the parametric method, we can now make use of each observation, not just the observation in the ACIS-I configuration. However, there are a few additional complications with modeling these data, as compared to the simulations. First, the source emission includes an additional contribution from unresolved X-ray binaries (XRBs). X-ray binaries have a different, harder, spectrum than hot gas, and in theory the two could be distinguished even in a spatial analysis by their hardness ratios. For this analysis, since we only examine the 0.5-2.0 keV energy band, we treat the XRBs spatially. We assume the XRBs are distributed like the K-band light is distributed, and we resample the K-band image of the galaxy from 2MASS  into the same pixel coordinates as each Chandra image. Using two different methods, H11 estimated the unresolved XRB luminosity to be $2.95\times10^{40}$ erg s$^{-1}$ and $2.8\times10^{40}$ (significant figures quoted as listed in H11) in the 0.5-7.0 keV band. We use a value of $2.9\times10^{40}$ erg $s^{-1}$ (converted into the 0.5-2.0 keV band, assuming an absorbed $\Gamma=1.6$ powerlaw) for the total luminosity of the XRB component.

Second, we found that within the core of the galaxy, even after accounting for XRBs the profile does not quite match the flat profile predicted by the $\beta$-model. Since the goal of this method is to study the faint diffuse emission at large radii, for which the behavior of the core is unimportant, we mask out the central 0.7 kpc (12 pixels) of the galaxy and fix the core radius $r_0$ to this value in the fit.

Finally, we detect and mask point sources separately for each observation. The deeper observations therefore typically have more point sources resolved. The radii of the masked regions also vary between observations, since these radii depend on the size of the psf, which varies depending on the location of the aimpoint. In the future it should be possible to combine the observations into a single analysis, but that requires additional simulations and testing which may be explored in future work. Because of the different point source masks in each observation, the luminosity of unresolved XRBs will vary between observations as well, so we include a fourth parameter $f_{\text{XRB}}$ which defines the fraction of XRB emission that is not masked out (the total $L_{\text{XRB}}$ is assumed to be $2.9\times10^{40}$ erg $s^{-1}$).

We use \verb"emcee" again to do the MCMC computations. We use 50 walkers, and for each walker discard the first 1000 elements of the chain before collecting the next 100 for analysis. The pdfs for the four fit parameters are displayed below (Figure 11), for all six observations. We also include the pdfs for $\beta$ and $N$, as estimated from H11 and H06. Since each observation has a different duration and configuration, we convert $N$ for each observation into the ``effective'' $N$ that would be measured if the source were positioned as in observation 11868 (the ACIS-I off-axis observation) To do this, we generate exposure maps for thermal $0.5$ keV plasma emission in the 0.5-2.0 keV energy band for each observation, and use the ratios of the exposure maps for the conversion.

\begin{figure*}
     \begin{center}
            \includegraphics[width=1.0\textwidth]{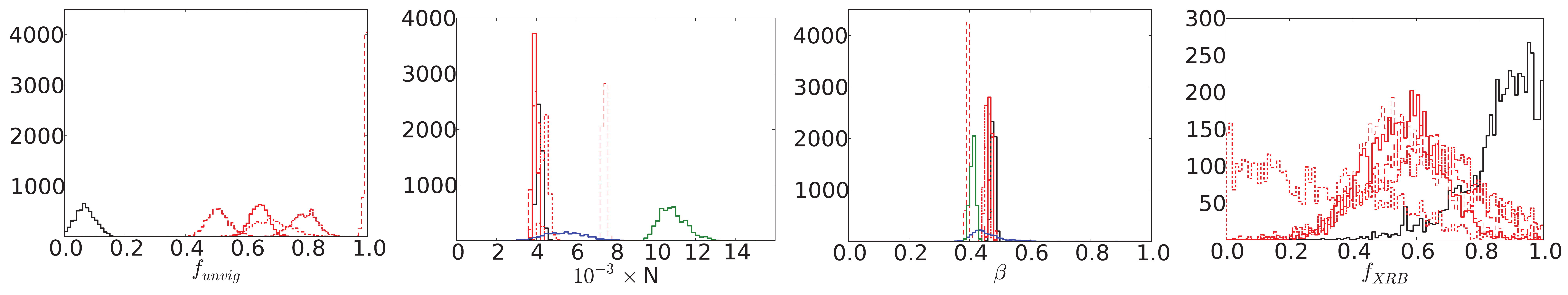}
    \caption{\small Posterior probability distribution functions (pdfs) for the model parameters in our parametric fits to six observations of NGC 720. The four model parameters are $f_{\text{unvig}}$ (the fraction of the background counts attributed to the unvignetted background), N (the number of source counts in the model, within a radius of 400 pixels), $\beta$ (the slope of the surface brightness profile), and $f_{\text{XRB}}$ (the fraction of the total XRB emission which is not masked out by the point source detection algorithms). In each plot, the black curve represents obs id 11868, the observation taken with the ACIS-I configuration, and the red curves represent observations taken in the ACIS-S configuration (solid = 7372, dotted = 8448, dot-dashed = 8449, dashed = 7062, thin dashed = 492). For N and $\beta$, we also present the pdfs for these parameters as estimated from our fits to the profiles of H06 (blue) and H11 (green), discussed in the text. There is strong convergence between all of our model fits, with the exception of obs id 492 which is very significantly affected by flaring. Our results are also consistent with the results of H06 but with much smaller uncertainties. }
    \end{center}
\end{figure*}

Five of the six observations give essentially equivalent results for the source parameters $N$ and $\beta$, which is a strong sign of success for our method. The sixth observation, with obs id 492, is significantly affected by flaring and the overall count rate is significantly elevated compared to the other observations. The effect of this high background is evident from the best-fit value of $f_{\text{unvig}} = 0.99 \pm0.01$. If other observations were not available and we needed to extract a more reliable measurement from this observation, we could filter the lightcurve more stringently, as did H06, in order to remove more of the contamination from flaring at the cost of reducing the amount of useable time. For this study, we just present the results as-is, and note that the flaring makes this observation unreliable. As we will show, this observation is an obvious outlier compared to the other five observations. 

The other observations cover a wide range of $f_{\text{unvig}}$, showing the need to fit for this parameter in each individual observation Observation 8448 is much shorter than the other observations, and so it has the least certain determination of $f_{\text{unvig}}$. We find higher values of $f_{\text{unvig}}$ for the observations in the ACIS-S configuration, which is likely related to the better sensitivity of the ACIS-S BI chips to low-energy instrumental X-rays. None of the observations yield very tight constraints on $f_{\text{XRB}}$, suggesting that broad constraints on the unresolved XRBs are sufficient for galaxies with hot gas as luminous as NGC 720. 

In Figure 12 we compare the results from the parametric and non-parametric methods for this galaxy. We take the pdfs from Figure 11 and generate surface brightness profiles for the source emission (gas + XRBs) from the model parameters, and we compare these profiles to the surface brightness profile as inferred from the non-parametric method for obs id 11868. Emission from the central 0.7 kpc was excluded from the parametric fit, although the non-parametric data includes this emission, so the leftmost nonparametric data point falls above the parametric fits. Beyond $\sim 3$' (23 kpc) the profile seems to steepen, suggesting that a second, steeper, component might improve the fit in the outer regions where the signal is below $\sim 1/5$ of the background.

At about 35 kpc the surface brightness profile becomes consistent with zero, although this radius is somewhat dependent on the choice of binning. Fitting a smoothing spline to the profile shows the average behavior at larger radii, and the spline remains above zero until about 50 kpc, where systematic uncertainties (probably a slight underestimate of $f_{\text{unvig}}$) take over the profile. For comparison, the outermost bin in H11 is 40-60 kpc, in which the claimed signal is at least $3\sigma$. This could be due to the deeper effective integration time in H11, since they analyze six observations (observation 492 is excluded) simultaneously, while the spline in Figure 12 is just fit to the results from a single observation. On the other hand, H06 has a $2.5\sigma$ detection of emission in their 62-90 kpc bin, which is difficult to reconcile with our results. One potential issue is that H06  relied solely on observation 492, and the background in this observation is particularly poorly-behaved due to all the flaring events throughout the integration. H11 (in their section 2.1.3) has some discussion of the effect of the deprojection procedure on the outermost annulus in H06 as well.

\begin{figure*}
     \begin{center}
            \includegraphics[width=1.0\textwidth]{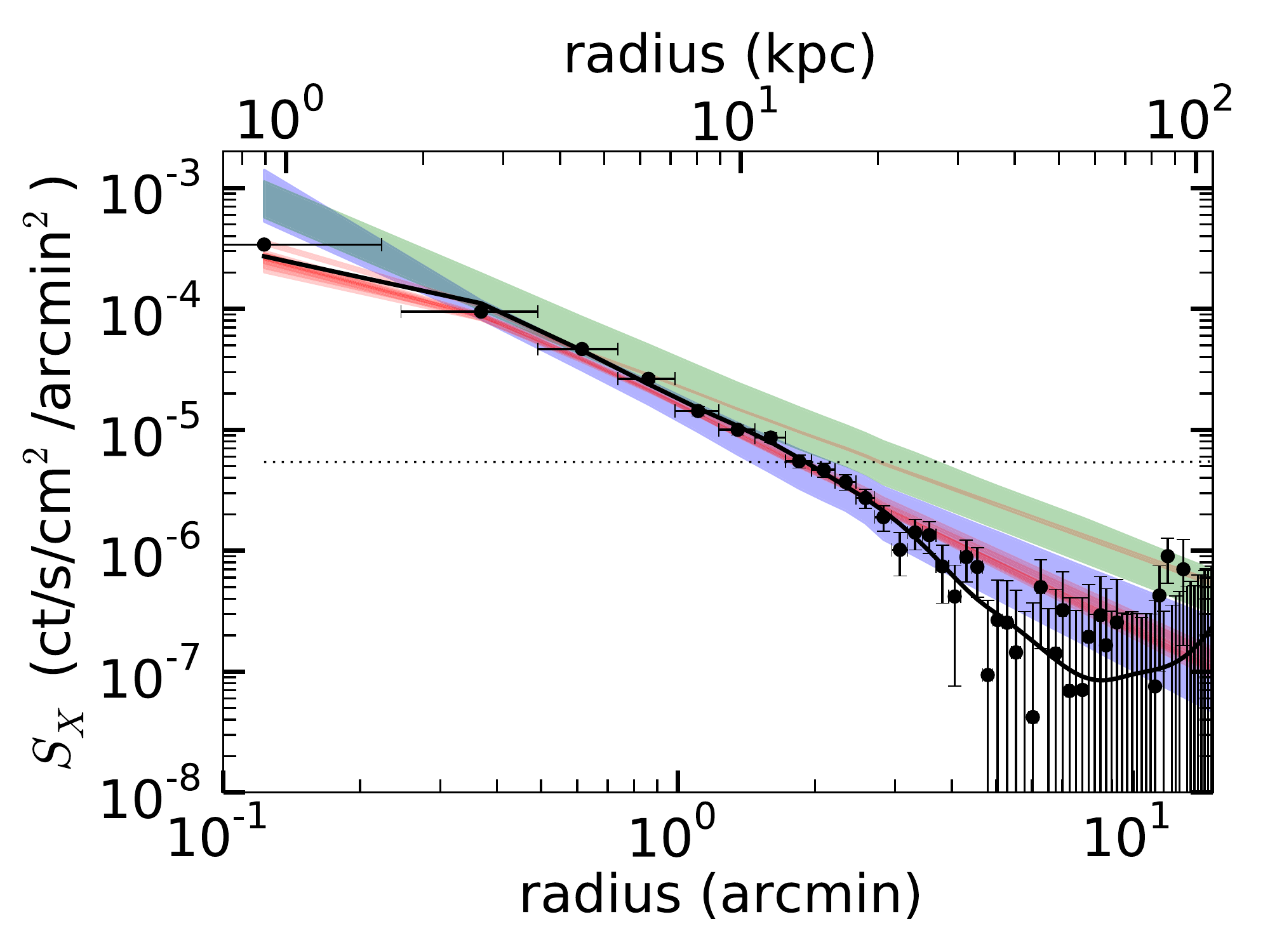}
    \caption{\small Comparison of background-subtracted surface brightness profiles for NGC 720 from the non-parametric method (data points) and parametric method (shaded regions). The red shaded regions correspond to the 68\% confidence regions for each of the six observations of NGC 720 (the outlier is observation 492 which is significantly affected by flaring).  The normalizations have been rescaled into the units of observation 11868 as discussed in the text. Blue and green shaded regions correspond to the 68\% confidence regions inferred from H06 and H11 respectively; H06 provides a reasonable fit to the data but H11 significantly overestimates the surface brightness profile. Since they employed the spectral method, both H06 and H11 have larger uncertainties than would be warranted by a fit to the surface brightness profile. The dotted black line shows the level of the background in observation 11868, as determined with the non-parametric method. The solid black line is a smoothing spline fit to the black data points; note that the profile appears to steepen faster than allowed by a $\beta$-model at $r\gapprox 25$ kpc.   }
    \end{center}
\end{figure*}

In this figure we also present the surface brightness profiles inferred from the density profiles of H06 and H11 (see section 5). In order to convert the density profiles into 0.5-2.0 keV X-ray surface brightnesses, we assume values of $Z = 0.6 Z_{\odot}$ and $kT = 0.5$ keV for the hot gas. Our choice of the $\beta$-model already assumes the temperature and abundance profiles are spatially uniform as well; this assumption is not quite true in detail, as H06 and H11 show, but their measured deviations from uniformity do not dramatically affect the shape of the surface brightness profile. We add an X-ray binary component to the surface brightness profiles as well, with the same range of luminosities as that of our fit to observation 11868, although the uncertainties in the gas density profile from H06 and H11 dominate over the XRB uncertainties. 

The H11 profiles are noticeably discrepant with the data. They are much closer to the anomalous fits to obs id 492, although H11 discarded this observation from their sample due to its flaring contamination. H06, which is actually fit to the data from observation 492 (after stringent flaring correction), provides a much better prediction for the surface brightness profile. The uncertainties in the surface brightness profiles from H06 and H11 are both significantly larger than our uncertainty, and seem larger than are warranted by the data. This can be explained by the small number of annuli used to derive the density profile, since H06 and H11 derive the density in each annulus from the X-ray spectrum. With only a few radii at which the density is measured, the range of profiles which can fit the data is much larger than the range of profiles which are consistent with the (much better spatially resolved) surface brightness data. 

We also compute the mass enclosed within each profile as a function of radius (Figure 13). In order to do this, we again assume values of $Z = 0.6 Z_{\odot}$ and $kT = 0.5$ keV for the hot gas, as well as assuming that these parameters are spatially uniform. We indicate with solid colors the region within which each model claims to measure the mass profile, and then we extrapolate the mass model out to 300 kpc and indicate the $1\sigma$ uncertainties around the extrapolated model.  We can see that, as H11 claimed, their mass model does indeed predict the hot halo contains most, if not all, of the missing baryons from the galaxy within 300 kpc (which is close to $R_{200}$ for this galaxy). This is also about a factor of two greater than the prediction of H06, and the two fits are inconsistent by a little more than $1\sigma$ at radii $\gapprox 80$ kpc. Our model for the hot gas is consistent with the low end of the range inferred by H06, although our uncertainties are about 50\% smaller. It is surprising that H11 is inconsistent with both H06 and our model, at all radii. There is some convergence at smaller radii where the background is less important, which is what we would expect if the issue in H11 were confusion between the spectrally similar background and source emission in H11, but there is still a discrepancy at all radii. 

\begin{figure}
\includegraphics[width=0.5\textwidth]{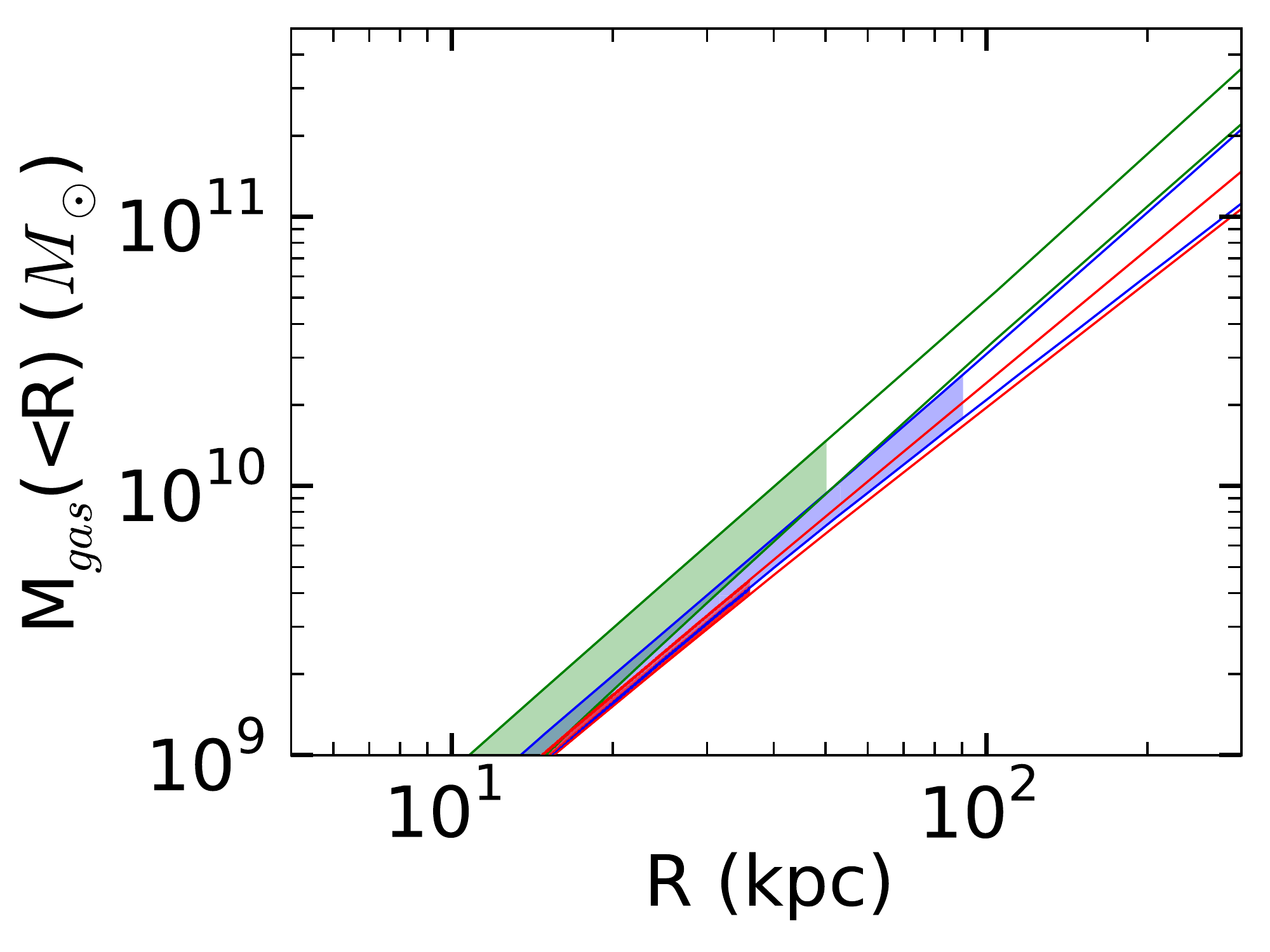}
\caption{$1\sigma$ regions corresponding to the estimated mass in the hot halo around NGC 720, from three different analyses of the galaxy. Blue corresponds to H06, green to H11, and red to our analysis. The regions are denoted with solid colors out to the radius of the outermost spectral bin (for green and blue), or at the outermost radius where the surface brightness is $1\sigma$ above the background (red); at larger radii the profiles are extrapolated outwards and the $1\sigma$ region is outlined. Discrepancies between H11 and the other profiles persist at all radii, even for the same assumed temperatures and metallicities for the hot gas.  Observation 492, which was significantly affected by flaring, has been excluded from this plot.  }
\end{figure} 

We note that the uncertainties here reflect only the uncertainties in the determination of the surface brightness profile. There are many additional uncertainties which bear on the determination of the mass of the hot halo as well, such as possible temperature and/or abundance gradients, gas clumping, multiphase effects, etc. The measured temperature and abundance gradients are fairly small, so the gas masses inferred within the dark shaded regions (where the gas is securely detected and measured) are fairly trustworthy. If the profile changes at larger radii (where most of the mass is inferred to lie), the true hot gas content at larger radii could be significantly different from the extrapolated value, so these values are much more uncertain and are illustrated solely to show the effects of the method of determining the profile on the inferred gas mass. The true mass in hot gas is unlikely
to be higher than our extrapolated values, though, since most plausible systematic errors (i.e. clumping or incomplete filling factor at large radii) reduce the total gas mass. Moreover, we note that the profile appears to steepen at $r \gapprox $ 3 arcmin in Figure 12, which suggests the density profile declines faster at large radii than our single-component $\beta$-model predicts. 

The results in Figure 13 have significant implications for the baryon budget of this galaxy. The mass of NGC 720 within $R_{200}$\footnote{H11 uses an alternative definition of the virial radius which puts $R_{vir}$ at about 400 kpc, but here we use the more common choice of $R_{200}$, which is about 300 kpc for this galaxy. The inferred mass budget looks the same for either choice.} is about $2.9\pm0.3\times10^{12} M_{\odot}$ \citep{Humphrey2011} , so multiplying by the Cosmic baryon fraction of 0.17 \citep{Hinshaw2013} yields an expected total baryonic mass of $5\times10^{11} M_{\odot}$. The stars comprise $1\times10^{11} M_{\odot}$ \citep{Humphrey2011}, leaving $4\times 10^{11} M_{\odot}$ of baryons missing from the budget. The extrapolated profile of H11 would therefore place most of these missing baryons in the hot gas, but we have argued that H11 overestimated the mass in hot gas, and the true mass is unlikely to be higher than about $1.5\times10^{11} M_{\odot}$. The ``missing'' baryons could lie in a cooler phase, such as the $\lapprox10^5$ K gas probed by \citet{Thom2012},  or outside the virial radius. We also cannot rule out the possibility of a second component in the million-degree gas, such as a low-metallicity, nearly uniform density diffuse medium, though we see no evidence for it in the surface brightness profile, and in fact the profile seems to steepen at $r\gapprox 25$ kpc instead of flattening as one would expect from such a model.

\section{Conclusion}

In this paper, we have accomplished several goals. We argued that neither spectral fitting nor spatial fitting, at their current levels of sophistication, are adequate for the analysis of extended faint X-ray emission below about 1/5 of the background. We have also argued that an ideal approach would study X-ray observations at the level of the events file, and therefore make use of the spectral and spatial information simultaneously. In order to take the first steps towards such an approach, we presented an improvement to spatial fitting, wherein we model the entire image within a single energy band (0.5-2.0 keV). This discards more detailed energy information but can be extended in this direction in future work.

We showed that a typical X-ray background can be decomposed into vignetted and unvignetted components. These components have different spectral shapes as well as different spatial distributions; in a given energy band, the different spatial distributions of these backgrounds can be used to constrain their relative contributions to an image. We introduced two methods of performing this decomposition. 

The {\it nonparametric} method excises a region around the source and fits the rest of the image in order to estimate the ratio of vignetted and unvignetted backgrounds, with no assumptions about the spatial distribution of the source emission other than that it must be localized to one region in the image which can be excluded. The nonparametric method will therefore not work for observations of diffuse emission which fills all or most of the field. 

The {\it parametric} method introduces a parametric model for the extended source (we examine $\beta$-models in this paper), and therefore is not limited to sources of small angular extent. We explored a number of possible likelihood functions for comparing the models to data. We decided upon a hybrid pixel-by-pixel likelihood function, with pixels binned where the psf is largest (and the correspond point source detectability the poorest). We showed that this likelihood function can reliably recover the shape of the background across the full spectrum from 100\% unvignetted to 100\% vignetted. This method works with either \verb"wavdetect" or with our manual point source detection algorithm, although we find the best results by masking point sources if either method detects them. 

We tested both methods for simulated extended sources observed with Chandra, in both the ACIS-I and ACIS-S configurations. We showed that we can recover the source emission well, recovering $\beta$ (a measure of the slope of the surface brightness profile) to 10\% accuracy and the total number of source counts to 25\% accuracy, for a source about half as bright as NGC 720. The method works better if the source is placed somewhat off-axis, so that its surface brightness profile can more easily be distinguished from the instrumental vignetting profile.

Finally we applied our method to the isolated elliptical galaxy NGC 720. Out of two previous studies of the density profile of its hot halo, one study (Humphrey et al. 2006) seems to predict an X-ray surface brightness profile that matches reasonably well with the observed data, while the other study (Humphrey et al. 2011) systematically over-predicts the X-ray surface brightness. Both studies only measure density or surface brightness in a handful of annuli ($8-11$), however, so the inferred surface brightness profiles have large uncertainties - much larger than are warranted by examination of the surface brightness profiles directly. We argue this is an inherent feature of the spectral method, and is one reason why combined spatial and spectral analysis is preferable. 

The surface brightness profile predicted by our models offers a much better fit to the data. With this method, we are also able to trace the source emission to well below a tenth of the background - a significant improvement over previous methods. Both our parametric and non-parametric methods largely agree with each other, although our parametric model is insufficiently detailed to capture deviations from the single-component $\beta$-model which we observe at radii $\gapprox3$'. 

The implied mass of the hot halo extrapolated to 300 kpc is $\lapprox 1.5\times10^{11} M_{\odot}$. This is insufficient by about a factor of three to bring the galaxy to baryonic closure, in contrast to the conclusions of \citet{Humphrey2011}. We instead estimate that, after accounting for the stars and the hot gas, NGC 720 is missing over half of its baryons. This brings the galaxy into closer agreement with our estimates in \citet{Anderson2013} of the amount of hot gas around typical L* galaxies based on stacked images from the ROSAT All-Sky Survey.

\section{Acknowledgements}
The authors would  like to thank the anonymous referee for a very helpful report that substantially improved the clarity and generality of the manuscript. The authors would also like to thank E. Bell, E. Churazov, A. Evrard, O. Gnedin, C. Miller, J. Miller, and M. Ruszkowski for helpful suggestions and comments on draft versions of the manuscript, as well as M. Miller and C. Slater for useful conversations relating to this work. J. Davis also kindly explained some of the features in MARX, and the documentation for all of the Chandra analysis software was uniformly excellent and informative. The authors would also like to thank T. Gaetz and P. Plucinsky for help interpreting the stowed Chandra background. Since the original data from H06 and H11 were unavailable, the free software WebPlotDigitizer (http://arohatgi.info/WebPlotDigitizer/), which is available under a GNU General Public License, was used to read the density profiles from these papers. This research has been funded by NASA ADAP grant NNX11AJ55G. This research has made use of data and/or software provided by the High Energy Astrophysics Science Archive Research Center (HEASARC), which is a service of the Astrophysics Science Division at NASA/GSFC and the High Energy Astrophysics Division of the Smithsonian Astrophysical Observatory. This research has made use of NASA's Astrophysics Data System. \\

\appendix

\counterwithin{figure}{section}

\section{Appendix 1: Nongaussianity of the Chandra psf}

The true Chandra psf is not exactly Gaussian. It is close to Gaussian in the core, but has wings which are more extended than a Gaussian. We can check how important these wings are by running  \verb"mkpsfmap"  with larger ecfs, using the 0.5-2.0 keV energy band. We compare the radii corresponding to higher ecfs to the expected radii if the emission were Gaussian with $\sigma$ equal to the radius enclosing 39.3\% of the emission. The results are below, in Figure A1, for ecfs of 75\%, 90\%, and 95\%. 

\begin{figure}
     \begin{center}
        \subfigure{%
            \includegraphics[width=0.33\textwidth]{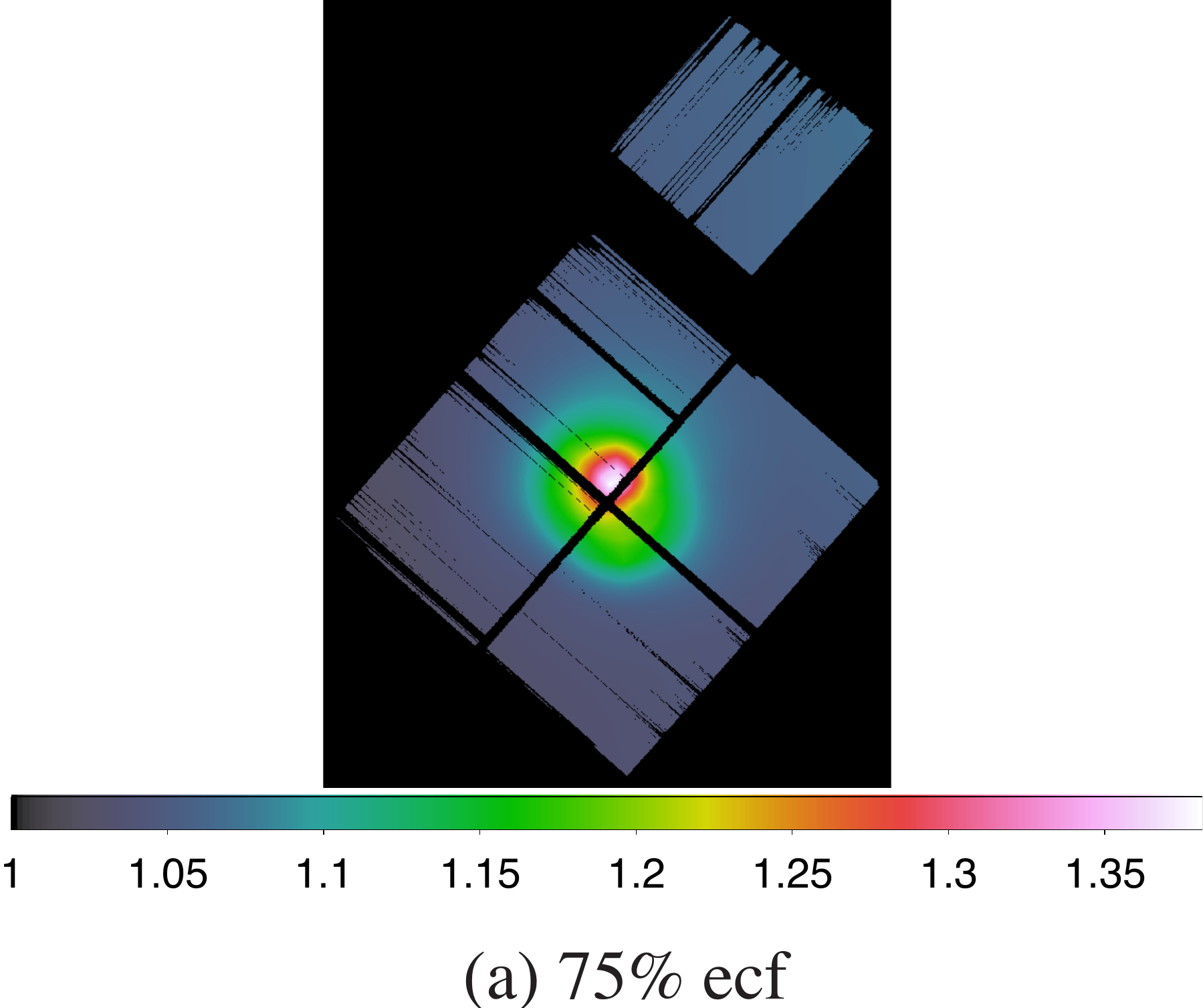}
        }%
        \subfigure{%
           \includegraphics[width=0.33\textwidth]{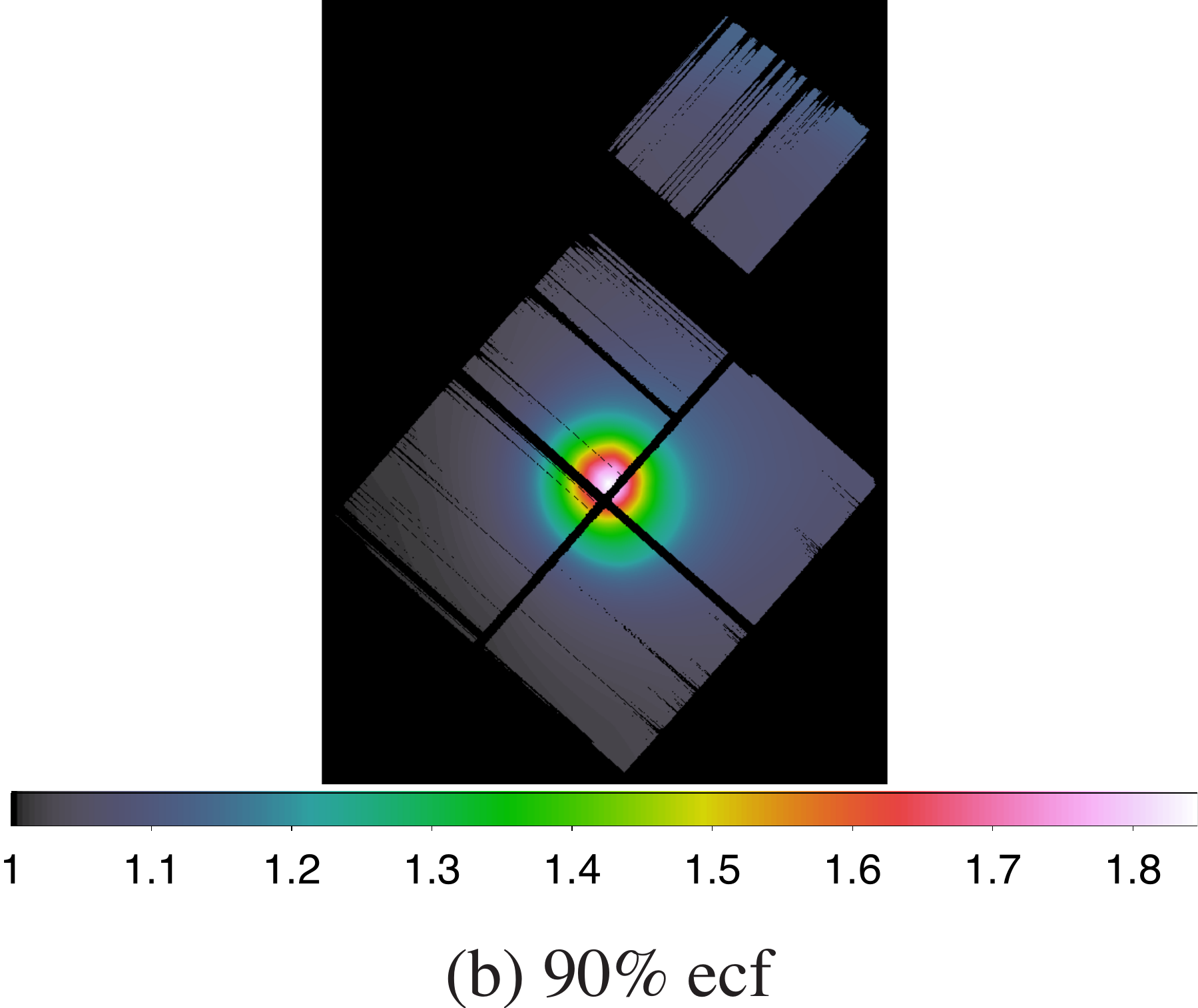}
        }%
        \subfigure{%
            \includegraphics[width=0.33\textwidth]{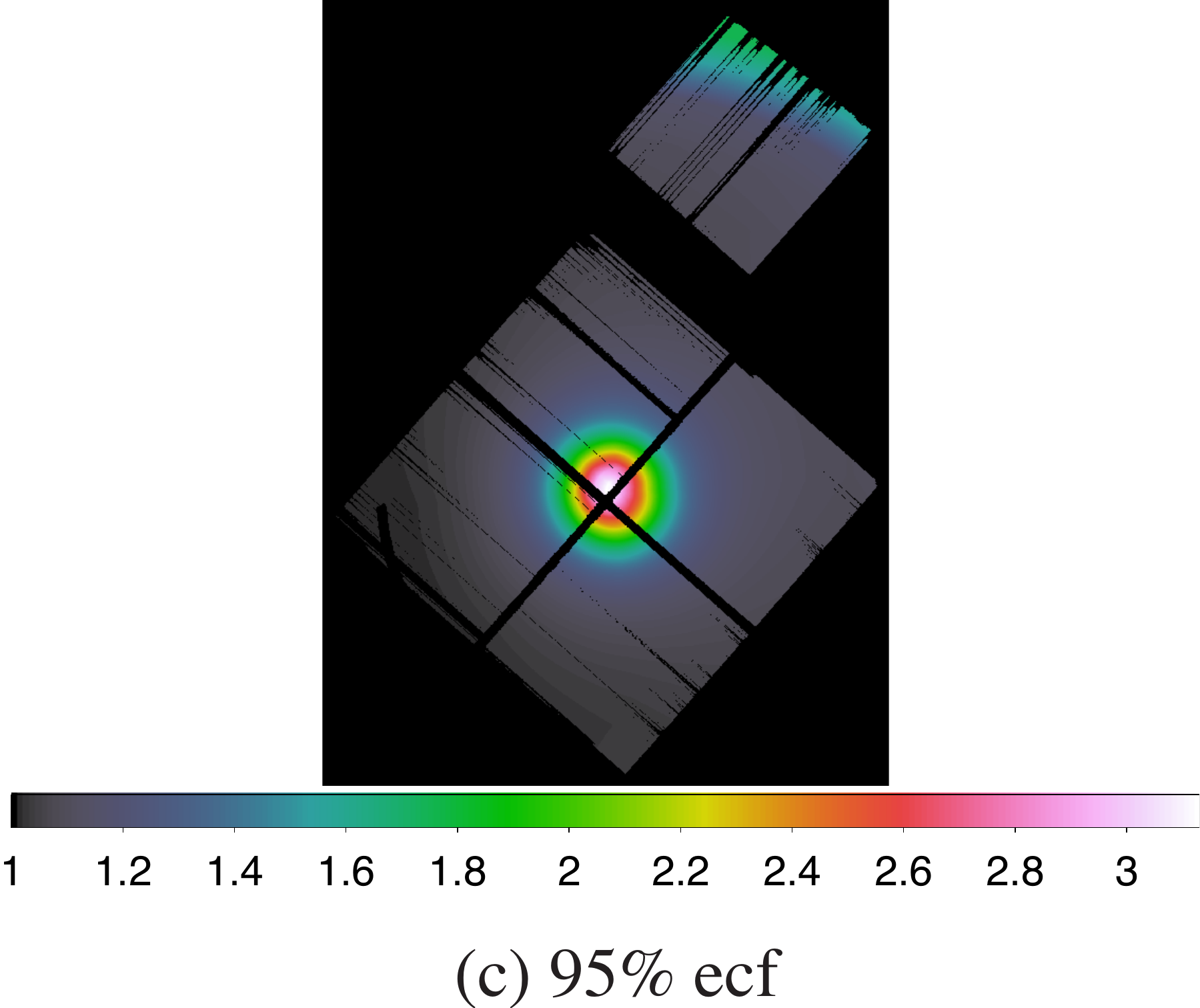}
        }%
    \end{center}
    \caption{\small Ratios of the radii corresponding to the (a) 75\% ecf, (b) 90\% ecf, and (c) 95\% ecf, each divided by the radius corresponding to the 39.3\% ecf. This entire ratio is then divided by the expected value of the ratio if the psf were perfectly Gaussian; thus the plots illustrate the amount of nongaussianity in the psf at the given ecf. There is little nongaussianity in the psf until radii corresponding to the 95\% ecf, suggesting that a Gaussian is a good approximation to the psf to about the 5\% level. There is a region in the center of the ACIS-I array where the psf is more strongly non-Gaussian, but the size of the psf here is small enough that deviations do not effect the image substantially. }
\end{figure}

Overall the psf does seem very close to Gaussian, with deviations becoming more visible at higher ecfs as the extended wings start to dominate. The deviations are not very significant across most of the detector even at an ecf of 95\%, suggesting that the deviations from Gaussianity are only at about the 5\% level. The psf is more non-Gaussian at the center of the detector, and a roughly 1-arcminute region near the very center is more extended than Gaussian even in the 75\% ecf comparison. Fortunately, the size of the psf in this region is still less than an arcsecond, so the effect of these nongaussianities is negligible, and we neglect them in this work.

\section{Appendix 2: Point Source Detection Algorithm Comparison}

Here we examine a 110 ks observation (obs id 8595) of the CDF-S. We reduce the image as described in section 3.  We then run \verb"wavdetect" on the image, masking out the $8\sigma$ ellipses around each point source. In Figure B1, we show the results from \verb"wavdetect" with our parameters, and compare them to the CDF-S 4 megasecond catalog of point sources \citep{Xue2011}. We do not recover all these point sources in the 110 ks image from observation 8595, which is unsurprising, but we do recover the bright point sources. There is one false positive in the image as well. Note that false positives are not a significant problem for us: they cost us a few background photons that are unnecessarily masked out, and probably bias downwards our estimate of the background flux rate slightly, but this effect is negligible for the low number of false positives generated by our significance threshhold. 

\begin{figure*}
\begin{center}
\includegraphics[width=10cm]{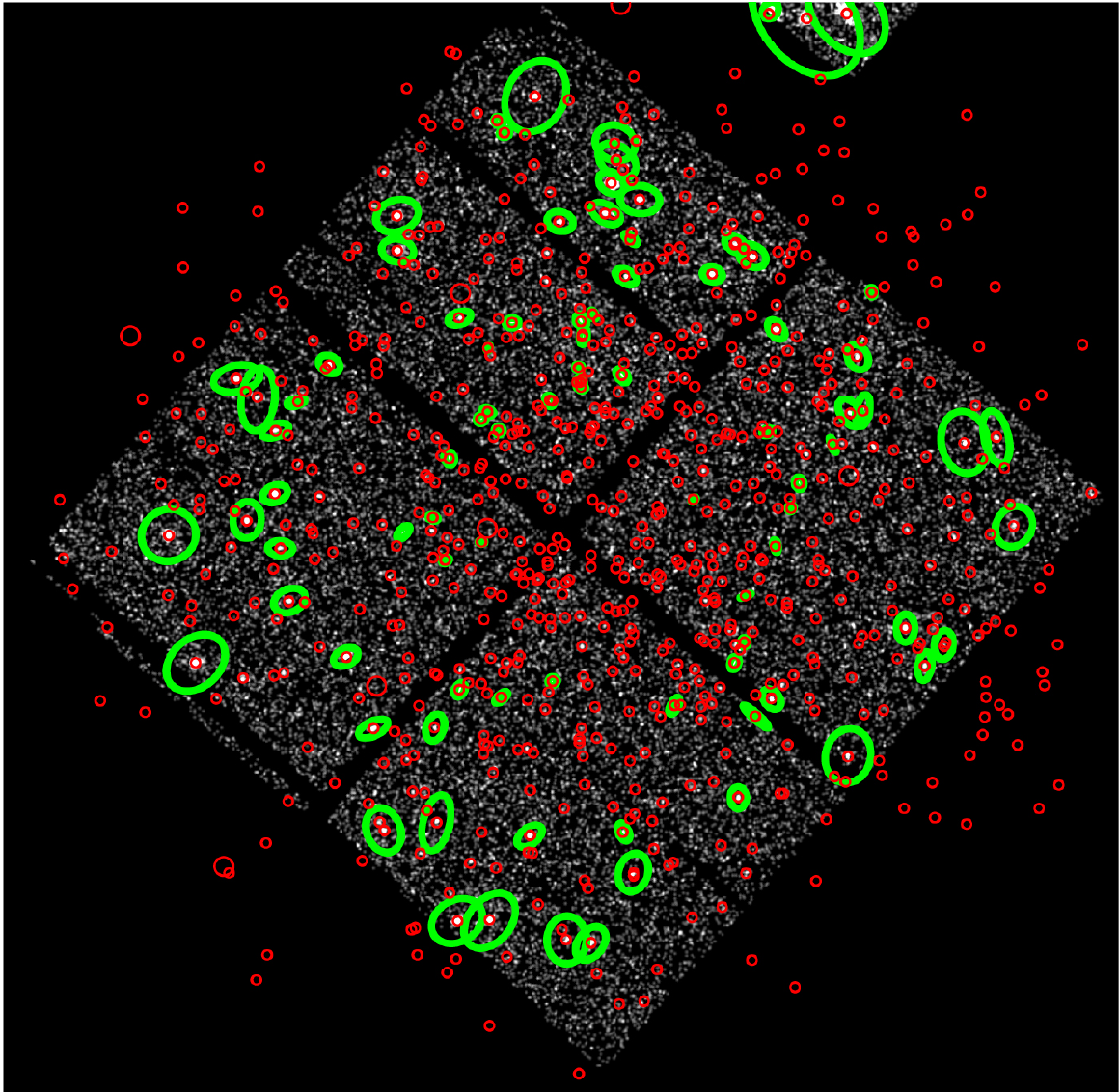}
\caption{Comparison of point sources in observation 8595 detected with {\it wavdetect} (green; note our unusually large ellipses) to all the known point sources in the field (red; for these sources the size of the circles is arbitrary) from the 4 megasecond analysis (Xue et al. 2011). We detect most of the bright point sources, and have one false positive (slightly to the left of the center of the ACIS-I array).}
\end{center}
\end{figure*} 

As a comparison, in Figure B2 we also illustrate the point sources detected in this image using our manual point source detection algorithm (section 3.3.1). This method measures the likelihood of obtaining the observed number of counts within a radius of the 90\% enclosed counts fraction (ecf) of the psf, at the location of every photon on the image. It is therefore more computationally intensive than \verb"wavdetect". It also detects slightly fewer point sources overall, but at large off-axis angles it appears to outperform  \verb"wavdetect"; this is notable in Figure B2 on the left side of the S2 chip where \verb"wavdetect" misses a fairly bright and extended source. None of the sources detected using the manual method are false positives. We compare the effects of the two different methods on recovering the correct unvignetted fraction of the background in Appendix 3. 

\begin{figure*}
\begin{center}
\includegraphics[width=10cm]{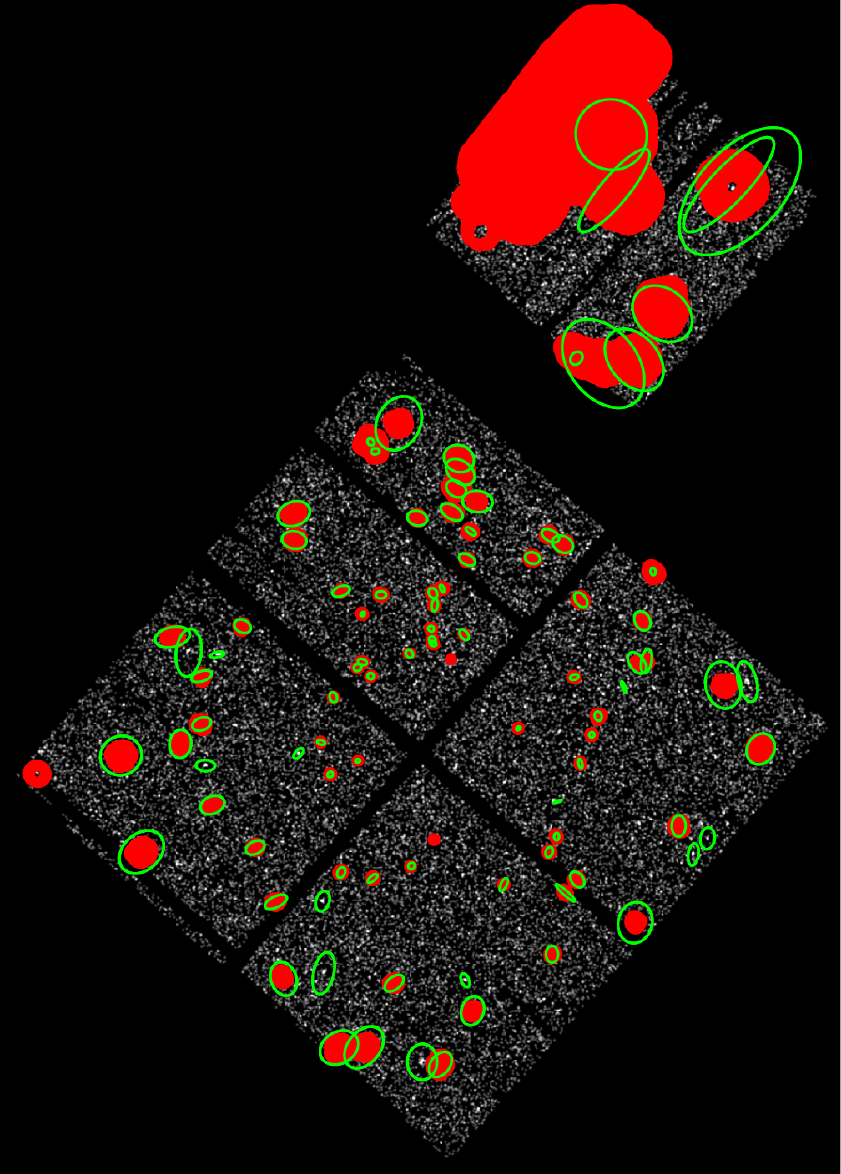}
\caption{Comparison of point sources detected with {\it wavdetect} (green; note our unusually large ellipses) and our manual point source detection algorithm (red). The two methods generally agree with one another, although each catches a few point sources missed by the other, and the manual method performs better on the ACIS-S chip very far off-axis. }
\end{center}
\end{figure*}

\section{Appendix 3: The Likelihood Function}

We explored a number of different likelihood functions for comparing observed and simulated images. A natural choice is the pixel-by-pixel function, defined as 

\begin{equation}L \equiv \prod_{\text{all pixels}} p(c_{\text{pix}} | m_{\text{pix}}) \end{equation}

where $p(c|m)$ is the Poisson probability of obtaining the observed number of counts $c$ in a pixel, given a model prediction $m$ for that pixel. We can also bin the data in various ways

\begin{equation}L \equiv \prod_{\text{all bins}} p(c_{\text{bin}} | m_{\text{bin}})  \end{equation}

where now we are adding pixels together within a bin before evaluating the likelihood. We explored binning by off-axis angle and by the size of the psf, in various combinations. We also included an additional component in the model to account for undetected point sources, which are more likely to be undetected at large radii. We discussed how to generate this component above, in section 3.3.2. 

To examine which of these likelihood functions gives the most reliable estimator, we simulated 11 images in each of the ACIS-I and ACIS-S configurations, based on the exposure maps for the CDF-S (obs id 8595, also examined in Appendix 2) and NGC 720 (obs id 7372) respectively. We used the 0.5-2.0 keV band and gave each image the same average flux per pixel as observation 8595 (i.e. 0.035 counts pix$^{-1}$ in the diffuse background). The 11 images used different values of $f_{\text{unvig}}$ ranging from 0\% to 100\%, in steps of 10\%. We also added simulated point sources to each image as described in section 3.3.2, and then applied our \verb"wavdetect" + manual point source masking algorithms as described in section 3.3.1.

For each image, we computed the likelihood as a function of $f_{\text{unvig}}$, using each of our different likelihood functions. We estimate uncertainties in the recovered value of the unvignetted fraction using the procedure discussed in \citet{Martin2008}. As long as the likelihood is well-behaved near the maximum, then the k$\sigma$ confidence interval around the maximum is ``bounded by the values that that correspond to the function $2ln(L)$ dropping by $k^2$.'' So the $1-\sigma$ region around the best-fit value of the unvignetted background fraction is just the region where the log-likelihood is within $0.61$ of the maximum value. We evaluate the effectiveness of recovering $f_{\text{unvig}}$ for each likelihood function by calculating the $\chi^2$, i.e. (recovered $f_{\text{unvig}}$ - true $f_{\text{unvig}}$)$^2$ / ($\sigma f_{\text{unvig}}$ (recovered)). This is not technically correct at the extreme values, since $f_{\text{unvig}}$ cannot go below 0 or above 1 and the errors are therefore not Gaussian at the edges. However, we checked the subset of simulations with the extreme points excluded (values of $f_{\text{unvig}}$ between 0.1 and 0.9) and obtained essentially the same results, so this effect does not appear to bias our results in any meaningful way. 

We consider the following likelihood functions: 

 - {\it pixel-by-pixel}, described in equation 1. 
 
 - {\it uniform radial bins}, where we divide the image into 20 annuli of the same radius, centered around the aimpoint. We bin together all the counts within each annulus before evaluating the likelihood.
 
 - {\it $r^{-1/2}$ radial bins}, where we divide the image into 20 annuli with radii decreasing as $r^{-1/2}$ (yielding approximately constant area per annulus), centered around the aimpoint. We bin together all the counts within each annulus before evaluating the likelihood.
 
- {\it psf size bins}, where we construct 20 bins based on the size of the at each pixel. Pixels with similar values for the radius of the 90\% ecf are binned together before evaluating the likelihood. 

- {\it hybrid}, which is a pixel-by-pixel likelihood function for the $N$\% of pixels where the psf is smallest, and we bin the remaining (100-$N$)\% pixels into $M$ bins determined by the size of the psf within that pixel. We explore three versions of hybrid function, with (N,M) = (10\%, 1 bin), (20\%, 2 bins), and (30\%, 3 bins).

We study each of these likelihood functions both with ({\it+ps}) and without ({\it-ps}) the correction for undetected point sources (section 3.3.2) included in the model. For reference, with 10 d.o.f., the fits from a likelihood function can be excluded at 95\% confidence if $\chi^2 > 18.31$ and at 99\% confidence if $\chi^2 > 23.21$.  In Table 1 we list the values of $\chi^2$ for each of the various likelihood functions we examined. To make the table easier to read, we have averaged the $\chi^2$ from ACIS-S and ACIS-I for each likelihood function. There were no large or systematic differences between the results in the two configurations, so no important information is lost by averaging the two values together. 

\begin{deluxetable}{lcccc}  % <--- column justification (center/left/right)
\tabletypesize{\footnotesize}
\tablecolumns{5}
\tablecaption{Likelihood Function Comparison}
\tablehead{   % column headings
  \colhead{Likelihood function} &
  \colhead{$\chi^2 $} &   \colhead{$\chi^2$ } &  \colhead{$\chi^2$ } &  \colhead{$\chi^2$ } \\
   \colhead{} & \colhead{ (no masking)} & \colhead{(manual method)} & \colhead{(wavdetect)} & \colhead{(manual \& wavdetect)}
  }
\startdata
pixel-by-pixel -ps & $>99$ & 30.6 & 31.4 & 25.6\\
pixel-by-pixel +ps & $>99$ & 19.6 & 14.3 & 16.7\\
uniform radial bins -ps & $>99$ & 24.0 & 26.5 & 16.9\\
uniform radial bins +ps & $>99$ & 35.1 & 20.0 & 21.2\\
$r^{-1/2}$ radial bins -ps & $>99$ &  24.6 & 50.6 & 18.2\\
$r^{-1/2}$ radial bins +ps & $>99$ & 61.2 & 22.0 & 22.0\\
psf size bins -ps & $>99$ & 22.1 & 35.3 & 23.4\\
psf size bins +ps & $>99$ & 35.6 & 20.0 & 26.1\\
hybrid, 1 bin -ps & $>99$ & $>99$ & $>99$ & $>99$\\
hybrid, 1 bin +ps & $>99$ &17.7 & 14.5 & 14.1\\
hybrid, 2 bins -ps & $>99$ & $>99$ & $>99$ & $>99$\\
hybrid, 2 bins +ps & $>99$ & 15.2 & 16.2 & 16.9\\
hybrid, 3 bins -ps & $>99$ & $>99$ & $>99$ & $>99$\\
hybrid, 3 bins +ps &  $>99$ & 22.8 & 19.4 & 24.1\\
\enddata
\tablecomments{\indent List of $\chi^2$ goodness of fit statistics for recovered values of $f_{\text{unvig}}$ using different likelihood functions and point source detection algorithms. Formally, the fits from a likelihood function can be excluded at 95\% confidence if $\chi^2 > 18.3$ and at 99\% confidence if $\chi^2 > 23.2$.  }
\end{deluxetable}

The wide variation in $\chi^2$ shows that the choice of likelihood function is important. The fit with the best average $\chi^2$ is the hybrid pixel-by-pixel method, with both \verb"wavedtect" and our manual point source masking method, and a component included in the model to account for undetected point sources. This method also gives reasonably-sized uncertainties in $f_{\text{unvig}}$ of a few percent for an observation of $\sim 100$ ks. We therefore adopt this method for most of the analysis in this paper.  \\

%\bibliographystyle{apj}
%\bibliography{paper.bib}{}

\end{document}